\documentclass[trackchanges,default]{aastex701}
\usepackage{latexsym,amsmath,amssymb}
\usepackage{graphics,graphicx}
\usepackage{natbib}
\usepackage{rotating}
\usepackage{enumitem}
\usepackage{lipsum}

\newcommand{\degm}{$^\circ$}

\def \rchisq {$\chi_{\nu} ^{2}$}

\newcommand{\rosat}{{\it ROSAT}}

\newcommand{\xmm}{{\it XMM-Newton}}

\newcommand{\swi}{{\it Swift}}
\newcommand{\cha}{{\it Chandra}}
\newcommand{\nustar}{\textit{NuSTAR}}

\newcommand{\msun}{$\rm M_{\odot}$}

\newcommand{\fluxcgs}{erg~s$^{-1}$~cm$^{-2}$}
\newcommand{\lumcgs}{erg~s$^{-1}$}
\newcommand{\chisq}{$\chi ^{2}$}

\begin{document}

\title{High Resolution X-ray Spectroscopy of the Nova-Like Cataclysmic Variable  BZ Cam using \cha\ HETG: Diagnosis of the ADAF-like (Advective) Hot Flow}

\author[0000-0001-6135-1144]{\c{S}\"olen Balman}
\altaffiliation{Corresponding Author}
\affiliation{Department of Astronomy and Space Sciences, Faculty of Science,  Istanbul University, Beyazit, 34119, Istanbul, Turkey}
\affiliation{Kadir Has University, Faculty of Engineering and Natural Sciences, Cibali 34083, Istanbul, Turkey}
\email[show]{solen.balman@gmail.com, solen.balman@istanbul.edu.tr}

\author[0000-0002-4162-8190]{Eric M. Schlegel}
\affiliation{Department of Physics and Astronomy, University of Texas-San Antonio, San Antonio, TX 78249, USA }
\email{eric.schlegel@utsa.edu }

\author[0000-0002-4806-5319]{Patrick Godon}
\affiliation{Department of Astrophysics \& Planetary Science, Villanova University, 800 Lancaster Avenue, Villanova, PA 19085, USA}
\affiliation{Henry A. Rowland Department of Physics \& Astronomy, Johns Hopkins University, Baltimore, MD 21218, USA}
\email{patrick.godon@villanova.edu }

\author[0000-0002-0210-2276]{Jeremy J. Drake}
\affiliation{Lockheed Martin Solar and Astrophysics Laboratory, 3251 Hanover St, Palo Alto, CA 94304, USA}
\email{jeremy.1.drake@lmco.com }

\author[0000-0003-4440-0551]{Edward M. Sion}
\affiliation{Department of Astrophysics \& Planetary Science, Villanova University, 800 Lancaster Avenue, Villanova, PA 19085, USA}
\email{edward.sion@villanova.edu }




\begin{abstract}

Nova-likes such as BZ Cam are high state Cataclysmic Variables showing hard X-ray emission that can be characterized with advective hot flows in the inner accretion disk. We explore \cha\ High Energy Transmission Grating (HETG) observations of BZ~Cam  for detailed line diagnosis and ionization conditions in the X-ray regime. We mostly find H- and He-like emission lines of Mg, Si, S, and Fe. All He-like line components of forbidden, intercombination and resonance lines are present. The R ratios of selected lines indicate plasma densities of a few $\times$10$^{12-14}$  cm$^{-3}$ and G ratios reveal temperatures (3-6)$\times$ 10$^6$ K where the Fe lines yield (1-3)$\times$ 10$^7$ K. The H to He line ratios and the R and G ratios show that the plasma is in a nonequilibrium  ionization condition, which is consistent with our previous X-ray results and the accretion flow in the X-ray region being an ADAF-like (advective) hot flow. Simultaneous fits of the HEG and MEG spectra or  the broadband joint spectra of \rosat, \cha\ zero order and \nustar\  yield temperatures 3.4-6.3 keV using a VNEI model of plasma emission (in XSPEC) or Bremsstrahlung emission. An additional power law is detected above 98\% Confidence Level in the broadband analysis. The orbital variations and the broadband spectra show dipping/veiling of the X-rays and  an additional warm absorber model with an ionization parameter log($\xi$) = 2.7 is required at the 3$\sigma$ level, along with the VNEI  model where  the HEG and MEG simultaneous fits yield the log($\xi$) = 3.6 .   
  
\end{abstract}



\section{Introduction}\label{sec:intro} 

Cataclysmic Variables (CVs) are close binaries in which the primary,
a white dwarf (WD), accretes matter from a late-type Roche-Lobe-filling main sequence star \citep{1995Warner}. CVs that are non-magnetic form an accretion disk reaching the WD surface. According to the standard accretion disk (SAD) theory \citep{1973Shakura}  half of the accretion luminosity originates from the disk and the other half from the boundary layer (BL) region very close to the WD \citep{1974Lynden-Bell}. SAD finds that for high accretion rate states ($\dot M_{acc}$$\ge$10$^{-(9-9.5)}$M$_{\odot}$), the BL should be optically thick and prominent in the soft X-rays and EUV (kT$\sim$10$^{(5-5.6)}$ K; \citep{1995Popham,1995Godon,2014Suleimanov,2015Hertfelder,2017Hertfelder}, whereas in the low accretion rate regime it should stay optically thin emitting hard X-rays \citep{1993Narayan,1999Popham}.
It has been shown that the optically thin BLs can be radially extended, advecting part of the energy to the WD as a result of their inability to cool \citep{1993Narayan}.
SAD is often found inadequate to model high state CVs in the UV, as well as some eclipsing quiescent dwarf nova and predicts a spectrum that is bluer than the observed UV  spectra indicating that a hot optically-thick inner flow including the BL is non-existent \citep{2007Puebla,2010Linnell,2017Godon}. As  a result, recent standard disk models have used a truncated inner disk \citep[e.g.,][]{2017Godon} that models the UV data more adequately.  Moreover, a parameterized model of viscous dissipation produced for the Nova-like (NL) CV IX~Vel fits the optical and UV data well \citep{2021Hubeny} and indicates a relatively flat vertical structure of temperature in the disk where most of the dissipation occurs  from the heated surface layers without a need for inner truncation. However, these models can not accommodate and explain the extent and/or emission characteristics of the X-ray region (e.g., hard X-rays).  The accretion flows in high state CVs are well explained in the context of radiatively inefficient advective hot flows (advection dominated accretion flow - ADAF-like flows) as opposed to standard optically thick accretion flows in the inner disk, clarifying complexities faced in the X-rays and other wavelengths \citep[see][]{2022Balman,2020Balman}.\ Shock formation in ADAFs around accreting WDs has been calculated \citep{2021Datta} and a shock can occur around $\sim$1.3$\times$10$^9$ cm from the WD surface which can explain the hard X-ray emission irrespective of the accretion rates.

The nonmagnetic NLs are found mostly in a state of high mass accretion rate (a few $\times$10$^{-8}$\msun\ yr$^{-1}$ to a few $\times$10$^{-9}$ \msun\ yr$^{-1}$). The VY Scl-type subclass exhibits high states and occasional low states of optical brightness while the UX UMa sub-type
remains in the high state \citep{1995Warner}; BZ Cam is a VY Scl-type system. 
All NLs show emission lines in the optical and/or UV wavelengths.
Bipolar outflows and/or rotationally dominated winds from NLs are
detected typically in the FUV through the P Cygni profiles of the resonance doublet of
C~IV \citep{1985Sion}. Mass loss rates of winds  are $\le$ 1\% of the accretion rate, with velocities of 200-5000 km s$^{-1}$ \citep{2004Kafka,2002Long}.

Observations of nonmagnetic quiescent CVs at low mass accretion rates
have yielded  hard X-ray spectra consistent with an optically thin
multi-temperature isobaric cooling flow model of plasma emission  (kT$_{max}$=6-50 keV) \citep[see][] {2005Pandel,2006Kuulkers,2011Balman,2020Balman,2025Balman}.
At high mass accretion rates ($\dot M_{acc}$$\ge$10$^{-9}$ M$_{\odot}$ yr$^{-1}$),
as opposed to expectations from standard steady-state accretion flow scenarios, observations of NLs
have always shown a hot optically thin hard X-ray source and data from all X-ray missions in the past and present have been modeled with MEKAL/APEC, double MEKAL/APEC models or multi-temperature plasma models with luminosities $\le$ a few $\times$10$^{32}$ \lumcgs \citep{1996vanTeeseling,2002Mauche,2004Pratt,2014Page,2014Zemko,2017Dobrotka,2022Balman}.

BZ Cam is a VY Scl-type NL system with a period of 221 min \citep{1996Patterson}. The binary system
has an inclination of about i = 12\degm–40\degm. GAIA archive\footnote{https://gea.esac.esa.int/archive}\ yields a distance of 374$\pm$3 pc using the
parallax measurement. The temperature of the WD is found to be 
40000–50000 K using Far-Ultraviolet Spectroscopic Explorer (FUSE) and IUE data \citep{2017Godon}. Winds from BZ Cam are detected in the FUV resonance lines with P Cygni profiles along with the Balmer and He I lines. They show a bipolar structure with unsteady and
continuously variable outflow of 3000–5000 km s$^{-1}$. Time variability of the wind structure is from 100 s to 3000 s in the optical and UV
indicating episodic behavior \citep{2013Honeycutt}.  
BZ Cam has a bow-shock nebula \citep{1987Krautter,1995Griffith}, which is not consistent with a
planetary nebula origin. It indicates prominent O III and H$\alpha$
emission with an  H$\alpha$ luminosity of 5.1$\times$10$^{30}$ erg s$^{-1}$ and OIII
luminosity of 9.1$\times$10$^{30}$ erg s$^{-1}$ \citep{2020Tappert}.

\subsection{Previous X-ray studies of BZ Cam}

The X-ray  investigations of 
the nonmagnetic NL, BZ Cam,
have been conducted using  \rosat\, \swi, and \nustar\ \citep{2014Balman,2022Balman}. 
BZ Cam has an X-ray spectrum best-fitted with a nonequilibrium (NEI) plasma emission 
model with kT$\sim$ 8-10 keV. No blackbody emission has been recovered with a 7 eV upper-limit for the blackbody effective temperature. In addition, there is some indication of a power law component with a photon index of 1.7-1.9 \citep{2022Balman}. This latter study finds the CEVMKL, a collisional equilibrium multi-temperature plasma model, inconsistent with the broad band spectrum in the 0.1-78.0 keV range, particularly around the 6-7 keV iron complex. 
The ratio (L$_{x}$/L$_{disk}$) (L$_{disk}$ from the UV-optical wavelengths)
yields considerable  inefficiency in the X-ray emitting region with about 0.1-0.5\% emission. As a result, the X-ray emitting
region in BZ Cam (and perhaps other NLs, see review by \citealt[][]{2025Balman}) is an extended optically thin hard X-ray emitting region with radiatively inefficient (ADAF-like) hot accretion flows in the inner disk in a fashion similar to X-ray binaries. Moreover, a study of the archival \cha\  HETGS data of CVs indicates line emission consistent with multi-temperature plasma in a nonequilibrium state \citep{2014Schlegel} that can be attributed to the characteristics of advective hot flows.  A P Cygni-type profile in the H-like iron line emission is detected from BZ Cam \citep{2022Balman} with the \nustar\ data that indicates fast collimated outflows in the X-rays (4500-8700 km s$^{-1}$). This is consistent with the nature of advective hot accretion flows as they can aid fast collimated outflows from disks because the flow has a positive Bernoulli parameter (the sum of the kinetic energy, potential energy and enthalpy). 

We proposed a \cha\ HETG observation to further our investigation of the accretion flows in the disk and improve our understanding of the X-ray emission from BZ Cam, and NLs in general. We expected that detailed X-ray spectroscopy using the ACIS-S HETG high resolution data should  reveal the complex line structures and ionization state and temperatures. Moreover, it should provide a better decomposition of all the emission and absorption components in the X-ray spectrum using broadband data in the 0.1-78.0 keV band (with joint analysis using archival \rosat\ and \nustar\ data). 

We also obtained HST COS data of BZ Cam simultaneously with the X-ray data in order to assess the mass accretion of the system during {\it Chandra} observations. 

This paper covers the analysis of our \cha\ HETG data and an assessment of our expectations and achievements using this 150 ks high resolution X-ray observation with the HETG. We characterize the nature and some properties of the ADAF-like accretion flows in BZ Cam using this data set. In our discussion, we elaborate on the  implications of
these results for the NL sub-class and for accreting WD binaries which will help to improve theoretical studies of disk modeling, MHD formalism, and perhaps shock formation. 

\section{Data and Observations}\label{sec:data}

BZ Cam was observed with the \cha\ Observatory \citep{2002Weisskopf}, High Energy Transmission Grating Spectrometer \citep[HETG][]{2005Canizares} in operation with the High Resolution Mirror Assembly (HRMA) utilizing a focal-plane imager, the  Advanced CCD Imaging Spectrometer, \citep[ACIS][]{2003Garmire}) as the readout CCD chip-set for the gratings for a  total of 154.5 ks on four separate pointings (PI=Balman) : 2024 September 23 (UT 07:36:13; 66.1 ks), 2024 September 24 (UT 14:03:23; 29.5 ks), and 2024 September 25 (UT 03:53:52; 29.5 ks), and 2024 September 25 (UT 20:04:41; 29.5 ks).  The HETG itself consists of two sets of gratings: 1) the Medium Energy Grating (MEG) that intercepts rays from the outer HRMA shells and is optimized for medium energies (0.4-5 keV; 31-2.5 \AA) 2) the High Energy Gratings (HEG) that intercepts rays from the two inner shells and is optimized for high energies (0.8-10 keV; 15-1.2 \AA). The High-Energy Transmission Grating Spectrometer (HETGS) operates from 0.4 to 10 keV and has a spectral resolution ($\lambda$/$\Delta\lambda$)=60-1000. 
The readout chip, ACIS-S, comprises a six CCD chip assembly  four of which are front-illuminated and two of which are back-illuminated CCDs. The ACIS-S3 has a moderate spectral resolution of E/$\Delta$E $\sim$10–30. The High Resolution Mirror Assembly (HRMA) of \cha\ produces images with a half-power diameter (HPD) of the point spread
function (PSF) of $<$ 0.$^{\prime\prime}$5.  The ACIS imager has a plate scale of 0.$^{\prime\prime}$49 per pixel.

Data were reprocessed using \cha\ Observatory data analysis software, CIAO 4.15-4.17 \citep{2006Fruscione}, with CALDB 4.11-4.12.
We followed the standard Chandra data analysis threads for analysis. 
We extracted the dispersed first-order and second order spectra using TGCat scripts \citep{2011Hue}, {\it tg$\_$create$\_$mask}, {\it tg$\_$resolve$\_$events} and {\it tgextract2} for both HEG and MEG instruments. We used standard pipeline-processed grating ancillary response functions ({\it arfs}) and grating redistribution matrix files ({\it rmfs}) at the highest resolution together with the spectra we produced using standard guidelines at the highest resolution of the HETG (both MEG and HEG). We note that rebinning spectra and using new generated {\it arf} and {\it rmf}  files at the given binning have been checked against the highest resolution spectra and have not been found to improve the spectral analysis, in general.
Further reductions and analyses of spectra and light curves were performed using HEASoft\footnote{https://heasarc.gsfc.nasa.gov/docs/software/heasoft} (version 6.32–6.35), using the XRONOS\footnote{https://heasarc.gsfc.nasa.gov/docs/software/xronos/xronos.html} or XSPEC\footnote{https://heasarc.gsfc.nasa.gov/docs/software/xspec/} software within the HEASoft distributions.

\section{Analysis and Results}\label{sec:ana}

\subsection{High Resolution Spectroscopy with HETG}\label{sec:hetg}

\subsubsection{Emission Lines detected using MEG and HEG}\label{sec:lhetg}

\begin{figure} 
\includegraphics[height=6cm,width=9.5cm]{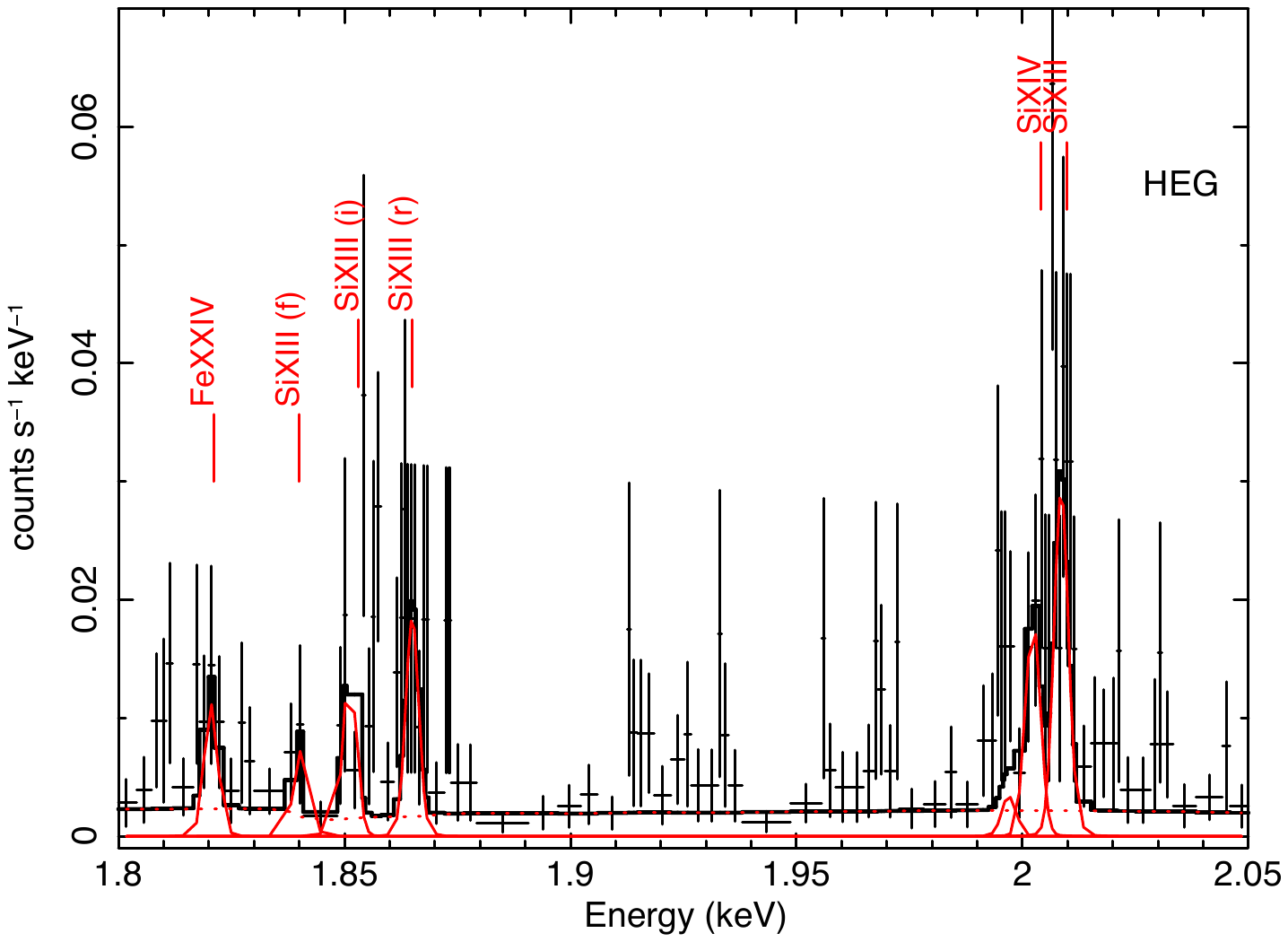} 
\hspace{-1.0cm}
\vspace{-0.7cm}
\includegraphics[height=6cm,width=9.5cm]{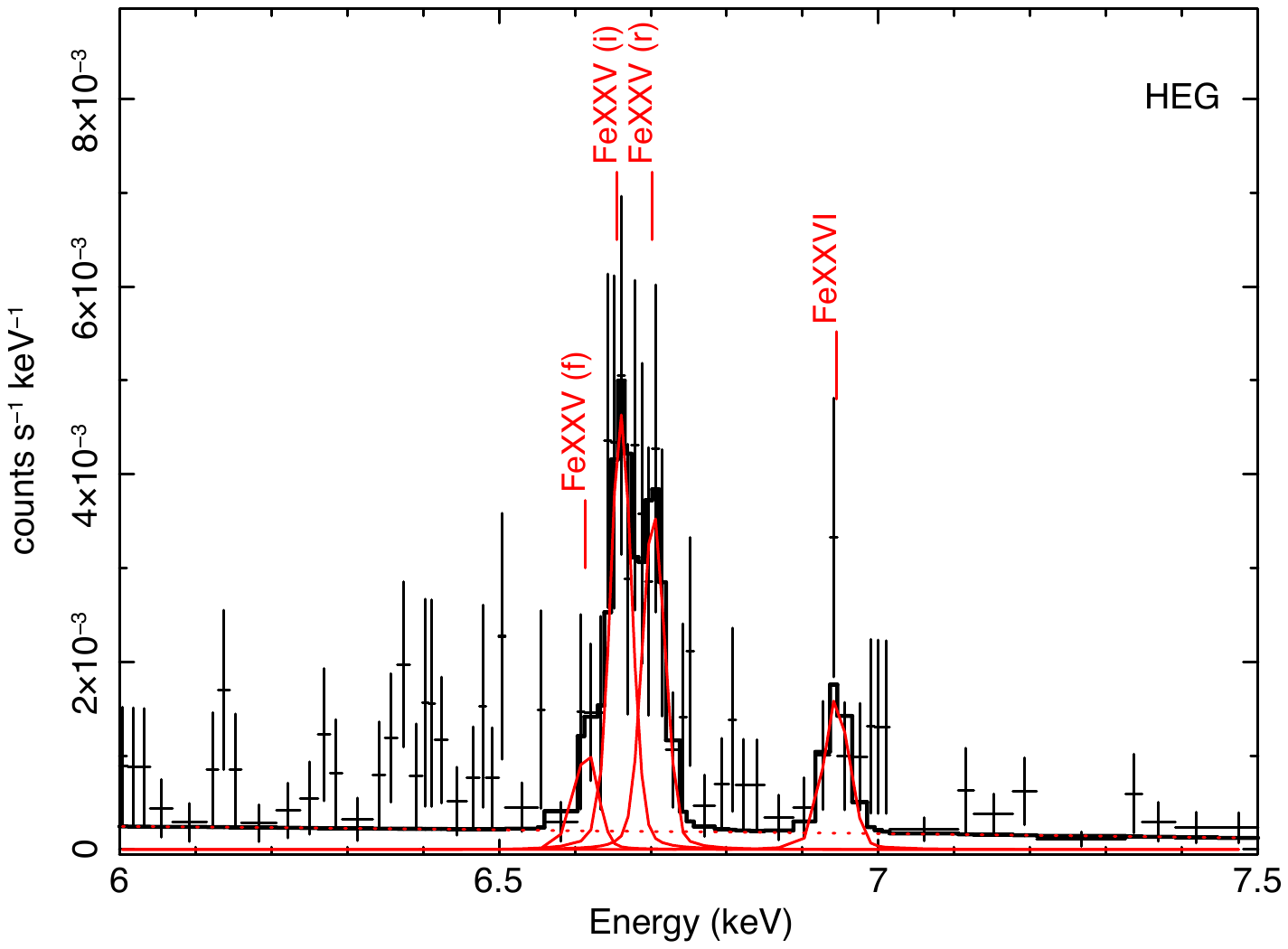}\\
\vspace{-0.7cm}
\includegraphics[height=6cm,width=9.5cm]{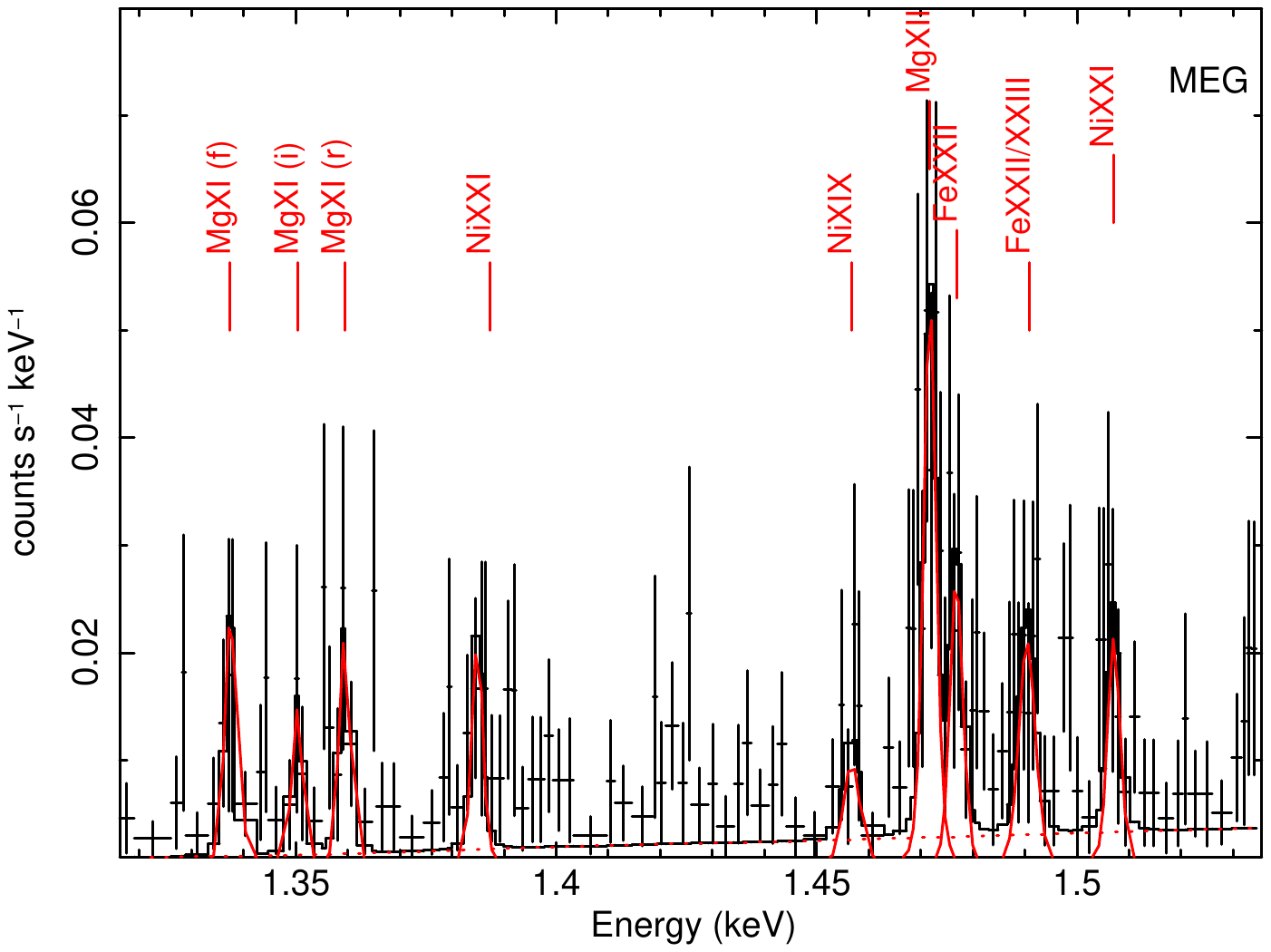}
\hspace{-1.0cm}
\includegraphics[height=6cm,width=9.5cm]{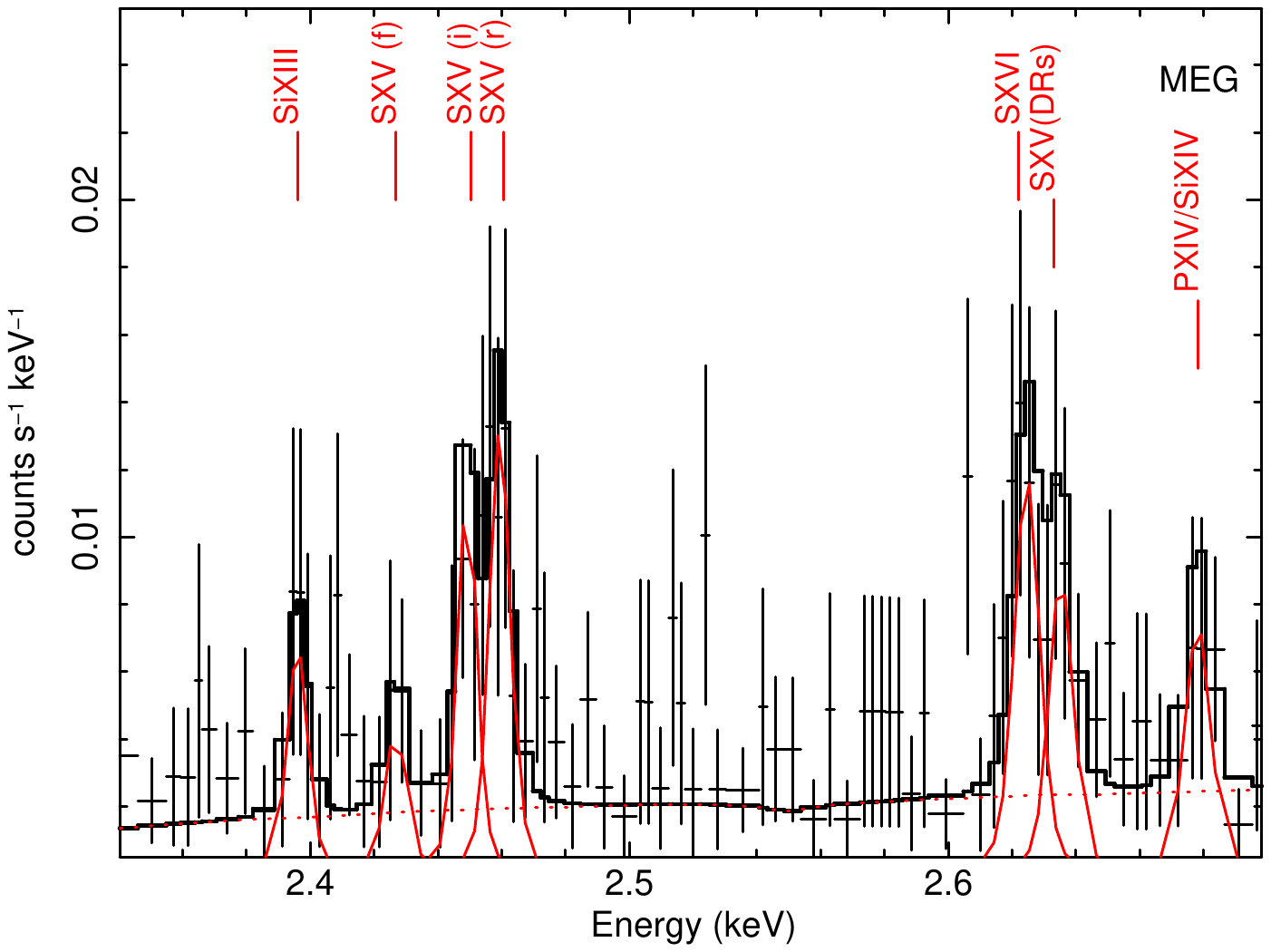}\\
\vspace{-0.5cm}
\includegraphics[height=6cm,width=9.5cm]{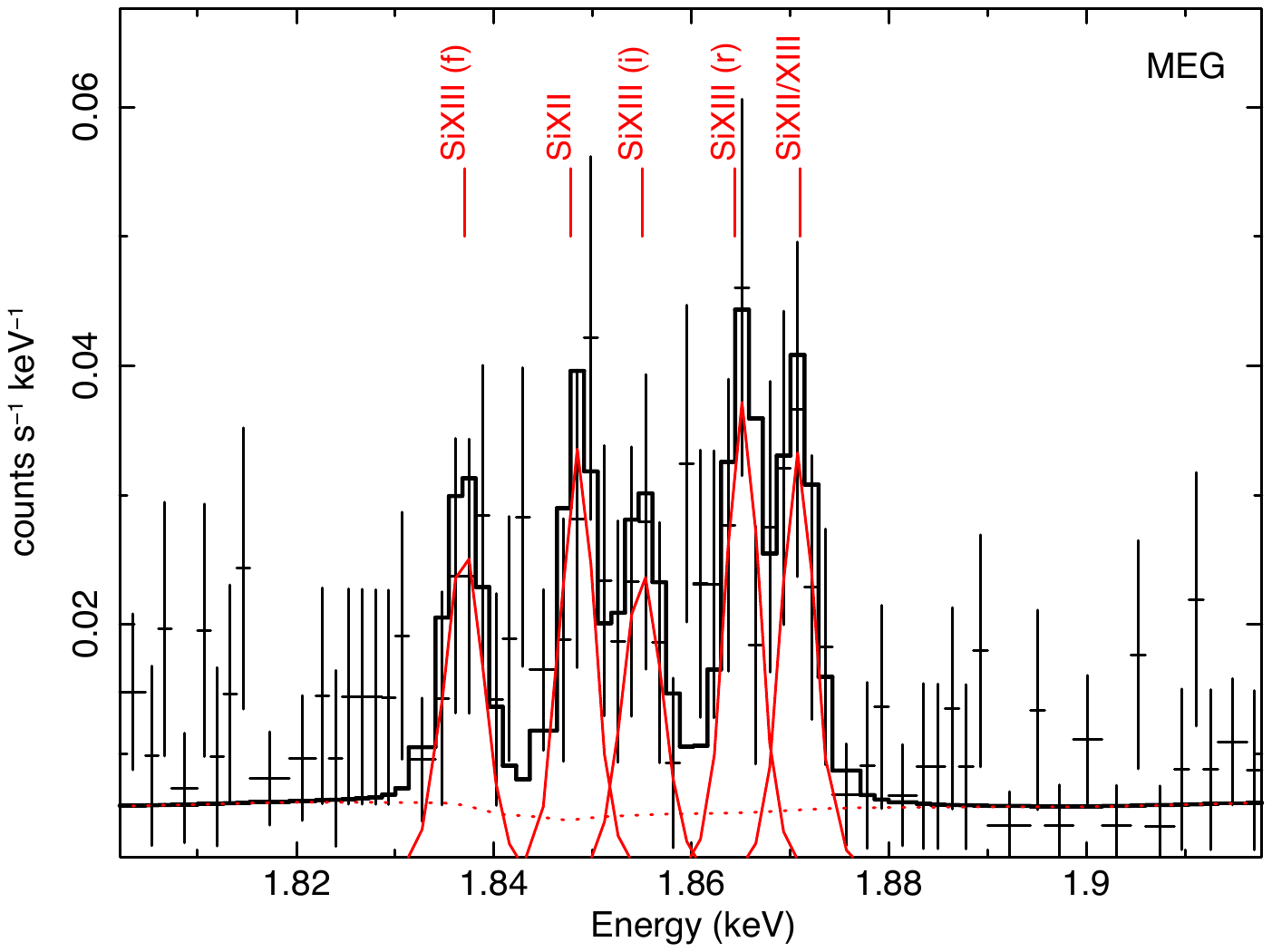}
\hspace{-1.0cm}
\includegraphics[height=6cm,width=9.5cm]{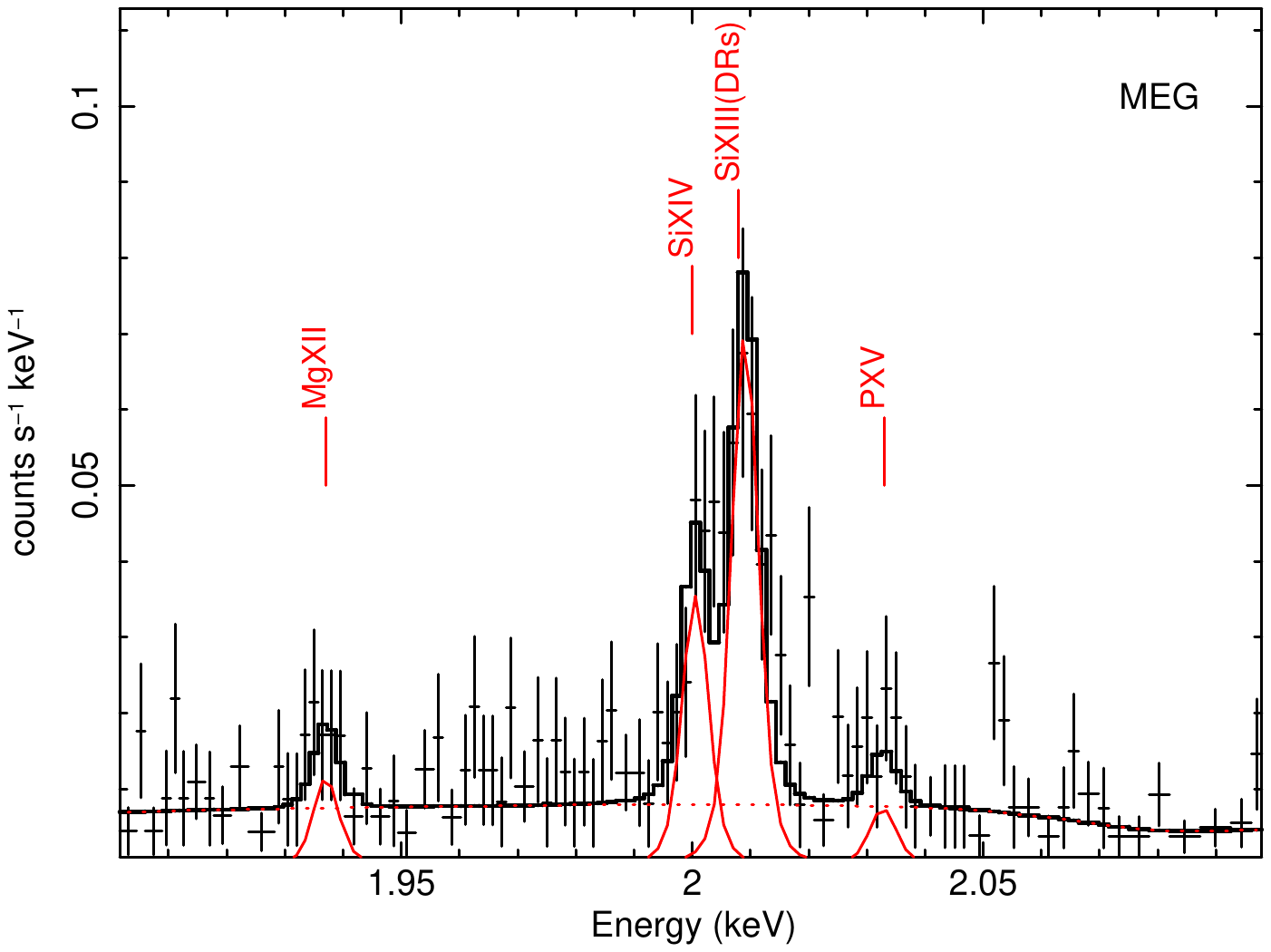}
\caption{
Top left and right hand panels show selected identified and fitted lines for the HEG spectrum of BZ Cam labeled individually (Silicon and Iron lines). Lines are fit with a GAUSS emission model and results are listed in  Table~\ref{tab:hlin}. The middle and bottom panels are of MEG spectra. The middle row displays selected, identified and fitted lines of H-like and He-like Magnesium (on the left) and H-like and He-like Sulfur lines along with some other detected lines (on the right). The bottom panels shows the Silicon line complex between 1.8 and 2.1 keV. Lines are fit with a GAUSS emission model and results are listed in  Table~\ref{tab:mlin}. 
\label{fig:lines}}
\end{figure}

One of the important directives of this work is determining the emission lines present in the X-ray spectrum of BZ Cam utilizing the high-resolution capacity of the \cha\ HETG. This should reveal the ionization conditions in the plasma and the nature of the accretion flows in the inner disk of the system. After creating spectra for given orders as described
in Sec.~\ref{sec:data}, we have combined the 1st order spectra, for MEG and HEG both, to increase the statistical quality in our analysis using the CIAO script {\it combine$\_$grating$\_$spectra} which yielded a count rate of 0.022 c s$^{-1}$ for MEG (0.4-5 keV; 31-2.5 \AA) and 0.014 c s$^{-1}$ for the HEG (0.8-10 keV; 15-1.2 \AA). We have used the {\it dmgroup} task to group the HEG and MEG spectra separately via choosing different group binning types like "ADAPTIVE", "ADAPTIVE$\_$SNR", or "SNR" where  SNR groups the energy bins with a given value of signal-to-noise ratio in the spectral bin which we chose mostly as 1-3. We have used different binning types to ensure the line detections and to see how they change with  binning-type values (we mostly used ADAPTIVE$\_$SNR=1.5 and SNR=1.5, which assumes a minimum of 1.5$\sigma$ in the spectral bins).
We have analyzed both the combined HEG and MEG  data separately and derived an emission line list for each instrument displayed in Table~\ref{tab:hlin} and Table~\ref{tab:mlin}, respectively.
We used the XSPEC task {\it identify} and the ATOMDB\footnote{http://atomdb.org} database  for line identifications. We recovered 24 lines with HEG and 35 lines with MEG.
The individual lines are fit via a Gaussian model or collections 
of two to three lines are fitted, if there is blending, along with a power law for the base/continuum around the line features. We note that the parameters of the power law for the base/continuum vary slightly across the fits as the plasma continuum changes in the given short/confined energy bands while fitting. We have  tabulated the lines with photon fluxes F$_{line}>$ 1$\times$10$^{-6}$ photons cm$^{-2}$ s$^{-1}$ ( $>$ 2$\times$10$^{-15}$  erg cm$^{-2}$ s$^{-1}$). Line tables denote the line centers in
kiloelectron volts and angstroms, together with the photon
fluxes of the lines derived from the normalization of the fits
and the corresponding energy flux. All errors are given at the
90\% Confidence Level for a single parameter.  Note that the errors on the energy fluxes are of the same percentile as the photon flux errors. Some of the detected important lines and their fits are displayed in Figure~\ref{fig:lines}. 

We have detected mostly emission lines around and above 1.0 keV out to the 6-7 keV iron line complex. The sensitivity of ACIS below 1.0 keV has degraded in time due to contamination buildup and BZ Cam is a dim source which precludes detection of significant signal below 1.0 keV. This range includes all the oxygen and neon lines (except the H-like neon emission which we did not detect). However,  we have recovered several Fe L-shell lines e.g., from Fe XVIII to Fe XXIV largely with MEG at a range of flux (1.6-9.7)$\times$10$^{-14}$ erg cm$^{-2}$ s$^{-1}$.

\begin{deluxetable}{lccll}
\tablewidth{0pt}
\tablecaption{The list of recovered emission lines in the HEG spectrum of BZ Cam. \label{tab:hlin}}
\tablehead{
Ion  &  Line Center $E_c$ & Line Center ${\lambda}_c$ & $K_{Gaussian}$  & Flux 
}
\startdata
 &   (keV) & (\AA) & 10$^{-6}$ & 10$^{-14}$ \\
 &   &  & Phot. cm$^{-2}$\ s$^{-1}$  & erg~cm$^{-2}$\ s$^{-1}$ \\
 \hline
Fe XXVI   &   $6.941^{+0.048}_{-0.019}$ & $1.786^{+0.005}_{-0.001}$ & $4.9^{+2.9}_{-2.9}$  &  5.4  \\
Fe XXV (r) &   $6.702^{+0.009}_{-0.009}$ & $1.850^{+0.003}_{-0.003}$  & $8.8^{+0.4}_{-3.8}$ & 9.4  \\
Fe XXV (i)   &  $6.655^{+0.009}_{-0.008}$ & $ 1.863^{+0.003}_{-0.003}$  & $12.0^{+0.4}_{-0.5}$ & 12.7  \\
Fe XXV (f)   &  $6.614^{+0.035}_{-0.028}$ & $ 1.875^{+0.008}_{-0.009}$  & $2.5^{+2.2}_{-0.9}$ & 2.7  \\
Ca XIX &  $3.875^{+0.003}_{-0.002}$ & $ 3.199^{+0.002}_{-0.002}$  & $3.5^{+2.3}_{-2.3}$ &  2.2 \\
Ar XVIII &  $3.316^{+0.018}_{-0.008}$ & $3.739^{+0.008}_{-0.020}$ & $2.6^{+1.7}_{-1.7}$  & 1.4  \\
Ar XVII/S XV Ly-$\beta$ & $3.138^{+0.006}_{-0.006}$ & $3.952^{+0.007}_{-0.007}$  & $2.8^{+1.9}_{-1.6}$ & 1.4  \\
S XVI  &    $2.621^{+0.006}_{-0.003}$ & $ 4.730^{+0.005}_{-0.010}$  & $2.7^{+1.4}_{-1.4}$ & 1.14  \\
 S XV (DRs) &  $2.584^{+0.006}_{-0.007}$ & $ 4.798^{+0.013}_{-0.011}$  & $4.3^{+2.2}_{-2.0}$ & 1.8  \\
 Si XIV &   $2.511^{+0.006}_{-0.005}$ & $4.938^{+0.009}_{-0.011}$  & $3.1^{+1.0}_{-1.5}$ & 1.25 \\
S XV (r) &  $2.461^{+0.002}_{-0.002}$ & $5.037^{+0.003}_{-0.003}$  & $2.5^{+2.3}_{-1.5}$ & 1.0  \\
S XV (i) &  $2.453^{+0.011}_{-0.006}$ & $5.054^{+0.011}_{-0.023}$ &  $3.0^{+2.0}_{-2.0}$ & 1.2 \\
S XV (f) &  $2.429^{+0.005}_{-0.005}$ & $5.105^{+0.011}_{-0.011}$ & $1.9^{+1.7}_{-1.7}$  & 0.74 \\
 Si XIII (DRs) &   $2.009^{+0.001}_{-0.001}$ & $6.169^{+0.004}_{-0.004}$ & $2.8^{+1.4}_{-1.4}$  & 0.9 \\
Si XIV &   $2.005^{+0.002}_{-0.002}$ & $6.184^{+0.005}_{-0.006}$ & $3.7^{+1.7}_{-1.7}$  & 1.2 \\
 Si XIII (DRs)  &  $1.997^{+0.007}_{-0.003}$ & $6.210^{+0.007}_{-0.022}$ & $1.9^{+1.2}_{-1.2}$ &  0.6 \\
Si XIII (r) &    $1.864^{+0.001}_{-0.001}$ & $6.652^{+0.004}_{-0.005}$ & $2.9^{+1.9}_{-1.9}$  & 0.9 \\
Si XIII (i) &    $1.854^{+0.004}_{-0.008}$ & $6.687^{+0.028}_{-0.091}$ & $2.2^{+1.6}_{-1.6}$ & 0.7 \\
Si XIII (f) &    $1.838^{+0.003}_{-0.004}$ & $6.745^{+0.015}_{-0.011}$ & $< 2.2$ & 0.6 \\
Fe XXIV  &  $1.821^{+0.005}_{-0.005}$ & $6.812^{+0.019}_{-0.019}$ & $1.3^{+0.9}_{-0.9}$ & 0.4 \\
Ni XXI &   $1.483^{+0.023}_{-0.023}$ & $8.360^{+0.131}_{-0.128}$  & $2.5^{+1.7}_{-1.7}$  & 0.6 \\
Mg XII &   $1.471^{+0.002}_{-0.001}$ & $8.426^{+0.007}_{-0.012}$ & $6.0^{+2.5}_{-2.5}$  & 1.4 \\
Ni XIX &   $1.455^{+0.003}_{-0.004}$ & $8.519^{+0.017}_{-0.022}$ & $2.4^{+1.7}_{-1.7}$  & 0.6 \\
Mg XI (r) &  $1.354^{+0.001}_{-0.004}$ & $9.155^{+0.031}_{-0.007}$ & $4.4^{+2.7}_{-2.7}$ & 0.96 \\ 
Mg XI (i) &  $1.340^{+0.028}_{-0.028}$ & $9.249^{+0.197}_{-0.189}$ & $< 3.7$ & $< 0.8$ \\ 
\hline
  &     &  & total line flux = &  50.3 \\
    &    &  w/o Fe XXV &  total line flux = &  25.5 \\
\enddata   
\tablecomments{
Gaussian line fits are performed using combined HEG spectra (order=1 (summed), see text for details) in the 0.8-10.0 keV (15-1.2 \AA) range.
A power law model is used for fitting the local continuum around the line of interest. DRs: Dielectric Recombination Satellite line.
$K_{Gauss}$ is the normalization for the Gaussian Line model. 
All lines are assumed to
be narrow as derived from the fits where the line widths $\sigma$ are constrained with the
spectral resolution of HETG and taken to be fixed at 0.001 keV.
All errors are calculated
at a 90\% confidence limit for a single parameter. Any upper limit is given at a 2$\sigma$
significance level.}
\end{deluxetable}

We have detected all the He-like and H-like lines of  Mg, Si, S, and Fe (all resonance (r), intercombination (i), and forbidden (i) emission lines of the He-like species are detected). HEG does not recover the forbidden line emission of Mg XI  (He-like Mg), but securely detects the H-like and He-like Fe (Fe XXVI and FeXXV) emission line complex. 
MEG recovers all these lines except the Fe complex since it is out of its sensitivity range. The strongest lines are found to be the He-like Fe XXV emission lines where the resonance line Fe XXV (r) is detected with a flux of 9.4$\times$10$^{-14}$ erg cm$^{-2}$ s$^{-1}$, the intercombination line Fe XXV (i) is detected at about 12.0$\times$10$^{-14}$ erg cm$^{-2}$ s$^{-1}$ and the forbidden line Fe XXV (f) is found to have a flux $\sim$2.7$\times$10$^{-14}$ erg cm$^{-2}$ s$^{-1}$. In general, the line flux ratios of the H-like emission lines to the He-like emission lines (r,i,f) have not been found to be more than about a factor of 2 (mostly around 1.0) . 
The ratio of  Fe XXVI line to the  intercombination or the resonance line of Fe XXV is 0.4-0.6. This value is, yet again, about 2 for the forbidden line ratio. 

An elaboration of these lines and diagnosis of the ionization condition, temperatures and densities from line  ratios is presented in the Discussion section \ref{sec:disc_hetg_diags}.

\begin{deluxetable}{lccll}
\tablewidth{0pt}
\tablecaption{The list of recovered emission lines in the MEG spectrum of BZ Cam. \label{tab:mlin}}
\tablehead{
Ion  &  Line Center $E_c$ & Line Center $\lambda_c$ & $K_{Gaussian}$  & Flux 
}
\startdata
 &   (keV) & (\AA) & 10$^{-6}$ & 10$^{-14}$ \\
 &   &  & Phot. cm$^{-2}$\ s$^{-1}$  & erg~cm$^{-2}$\ s$^{-1}$ \\
 \hline
Ca XX &   $4.410^{+0.005}_{-0.018}$ & $2.811^{+0.012}_{-0.003}$ & $2.5^{+1.7}_{-1.7}$  &  1.8  \\
Ca XVIII/XIX  &  $3.860^{+0.048}_{-0.024}$ & $ 3.212^{+0.020}_{-0.039}$ & $2.85^{+1.8}_{-1.8}$ & 1.8 \\
Ar XVIII &  $3.322^{+0.014}_{-0.013}$ & $3.733^{+0.014}_{-0.016}$ & $2.3^{+1.6}_{-1.6}$  & 1.2 \\
Ar XVII/S XV Ly-$\beta$ &    $3.122^{+0.001}_{-0.014}$ & $3.972^{+0.001}_{-0.019}$  & $4.9^{+2.0}_{-2.0}$ & 2.5 \\
 S XV/S XIV(DRs) &  $2.868^{+0.029}_{-0.025}$ & $4.323^{+0.043}_{-0.039}$  & $1.9^{+1.4}_{-1.4}$ & 0.9 \\
P XIV/Si XIV & $2.678^{+0.011}_{-0.003}$ & $4.629^{+0.005}_{-0.019}$  & $1.7^{+1.4}_{-1.4}$ & 0.54 \\
 S XIV/S XV(DRs)/Si XIV(CX) & $2.626^{+0.008}_{-0.004}$ & $4.722^{+0.007}_{-0.015}$  & $2.9^{+1.7}_{-1.7}$ & 1.2  \\
S XVI &    $2.620^{+0.002}_{-0.004}$ & $4.732^{+0.008}_{-0.003}$  & $2.0^{+1.8}_{-1.8}$ & 0.8 \\
S XV (r) &  $2.461^{+0.002}_{-0.003}$ & $5.037^{+0.005}_{-0.004}$  & $2.7^{+1.7}_{-1.7}$ & 1.1  \\
S XV (i) &  $2.450^{+0.004}_{-0.003}$ & $5.059^{+0.006}_{-0.009}$ &  $2.3^{+1.7}_{-1.7}$ & 0.9 \\
S XV (f) &  $2.427^{+0.005}_{-0.004}$ & $5.109^{+0.008}_{-0.010}$ & $1.4^{+1.3}_{-1.3}$  & 0.6 \\
 Si XIII/Si XIV Ly-$\beta$/S XIV(DRs) &  $2.396^{+0.013}_{-0.008}$ & $5.175^{+0.016}_{-0.028}$ & $2.5^{+1.6}_{-1.6}$  & 1.0 \\
P XV Ly-$\alpha$ &  $2.033^{+0.004}_{-0.006}$ & $6.099^{+0.019}_{-0.011}$ & $1.2^{+0.8}_{-0.8}$  & 0.5  \\
Si XIII(DRs)  &   $2.009^{+0.001}_{-0.001}$ & $6.172^{+0.002}_{-0.004}$ & $6.1^{+1.4}_{-1.4}$  & 2.0 \\
Si XIV &   $2.005^{+0.001}_{-0.001}$ & $6.184^{+0.004}_{-0.004}$ & $3.6^{+1.2}_{-1.2}$  & 1.2 \\
Mg XII  &  $1.937^{+0.002}_{-0.008}$ & $6.401^{+0.026}_{-0.005}$ & $1.1^{+0.7}_{-0.7}$ & 0.34 \\
 Si XIII/Si XII(DRs) &  $1.871^{+0.002}_{-0.008}$ & $6.627^{+0.028}_{-0.006}$ & $2.1^{+0.9}_{-0.9}$ & 0.6  \\
Si XIII (r)  &  $1.865^{+0.001}_{-0.001}$ & $6.647^{+0.002}_{-0.003}$ & $2.4^{+0.9}_{-0.9}$ & 0.7 \\
Si XIII (i) & $1.855^{+0.001}_{-0.001}$ & $6.684^{+0.002}_{-0.005}$ & $1.8^{+0.9}_{-0.9}$ & 0.53 \\
Si XII/Si XII (DRs)  & $1.848^{+0.003}_{-0.001}$ & $6.710^{+0.002}_{-0.010}$ & $2.3^{+1.0}_{-1.0}$ & 0.7 \\
Si XIII (f) &    $1.837^{+0.005}_{-0.004}$ & $6.749^{+0.014}_{-0.019}$ & $1.6^{+0.7}_{-0.7}$ & 0.5 \\
Ni XXI &   $1.507^{+0.001}_{-0.004}$ & $8.226^{+0.019}_{-0.006}$  & $1.2^{+0.7}_{-0.7}$  & 0.3 \\
Fe XXII/XXIII &   $1.490^{+0.002}_{-0.005}$ & $8.319^{+0.029}_{-0.010}$& $1.6^{+0.8}_{-0.8}$  & 0.4 \\
Fe XXII &   $1.477^{+0.001}_{-0.002}$ & $ 8.395^{+0.009}_{-0.005}$  & $1.6^{+0.8}_{-0.8}$ &  0.4 \\
Mg XII &   $1.472^{+0.001}_{-0.001}$ & $8.425^{+0.004}_{-0.005}$ & $3.2^{+1.1}_{-1.1}$  & 0.8 \\
Ni XIX &   $1.457^{+0.003}_{-0.004}$ & $8.511^{+0.022}_{-0.019}$ & $0.8^{+0.6}_{-0.6}$  & 0.2 \\
Ni XXI  &   $1.387^{+0.003}_{-0.003}$ & $8.937^{+0.015}_{-0.015}$ & $1.6^{+0.9}_{-0.9}$  & 0.35 \\
Mg XI (r) &  $1.359^{+0.004}_{-0.005}$ & $9.121^{+0.035}_{-0.025}$ & $2.0^{+1.1}_{-1.1}$ & 0.44 \\ 
Mg XI (i) &  $1.351^{+0.006}_{-0.003}$ & $9.181^{+0.018}_{-0.004}$ & $1.4^{+1.0}_{-1.0}$ & 0.3 \\ 
Mg XI (f) &  $1.337^{+0.001}_{-0.002}$ & $9.271^{+0.010}_{-0.006}$ & $2.5^{+1.4}_{-1.4}$ &  0.54 \\ 
Fe XX/XXI &   $1.275^{+0.015}_{-0.005}$ & $9.722^{+0.041}_{-0.112}$ & $2.6^{+1.1}_{-1.9}$  & 0.5 \\
Ni XXIII/XXIV &   $1.188^{+0.004}_{-0.003}$ & $10.436^{+0.029}_{-0.033}$ & $2.4^{+1.9}_{-1.9}$  & 0.45 \\
Fe XXIV &  $1.169^{+0.005}_{-0.005}$ & $10.599^{+0.047}_{-0.040}$ & $3.3^{+2.4}_{-2.4}$  & 0.6 \\
Fe XX/XXIV &  $1.128^{+0.003}_{-0.004}$ & $10.988^{+0.042}_{-0.024}$ & $9.7^{+4.8}_{-4.8}$ & 1.8 \\ 
Fe XIX  &  $1.110^{+0.001}_{-0.004}$ & $11.169^{+0.045}_{-0.007}$ & $4.0^{+3.4}_{-3.4}$ & 0.7 \\ 
Fe XVIII &  $1.083^{+0.001}_{-0.001}$ & $11.450^{+0.012}_{-0.004}$ & $4.8^{+3.7}_{-3.7}$ & 0.8 \\
Fe XVIII/XIX &  $0.871^{+0.010}_{-0.015}$ & $14.243^{+0.250}_{-0.160}$ & $2.6^{+2.2}_{-2.2}$ & 3.6 \\ 
\hline
  &    &  & total line flux = &  32.6 \\
\enddata   
\tablecomments{
Gaussian line fits are performed using combined MEG spectra (order=1, see text for details) in the 0.4-5.0 keV (31-2.5 \AA) range.
A power law model is used for fitting the local continuum around the line of interest. DRs: Dielectric Recombination Satellite line.
$K_{Gauss}$ is the normalization for the Gaussian Line model. 
All lines are assumed to
be narrow as derived from the fits where the line widths $\sigma$ are constrained with the
spectral resolution of HETG and taken to be fixed at 0.001 keV.
All errors are calculated
at a 90\% confidence limit for a single parameter. Any upperlimit is given at a 2$\sigma$
significance level.}
\end{deluxetable}

\subsubsection{Joint spectral analysis of HEG and MEG}\label{sec:jhetg}

\begin{figure*}
\includegraphics[height=7.cm,width=9.7cm,angle=0]{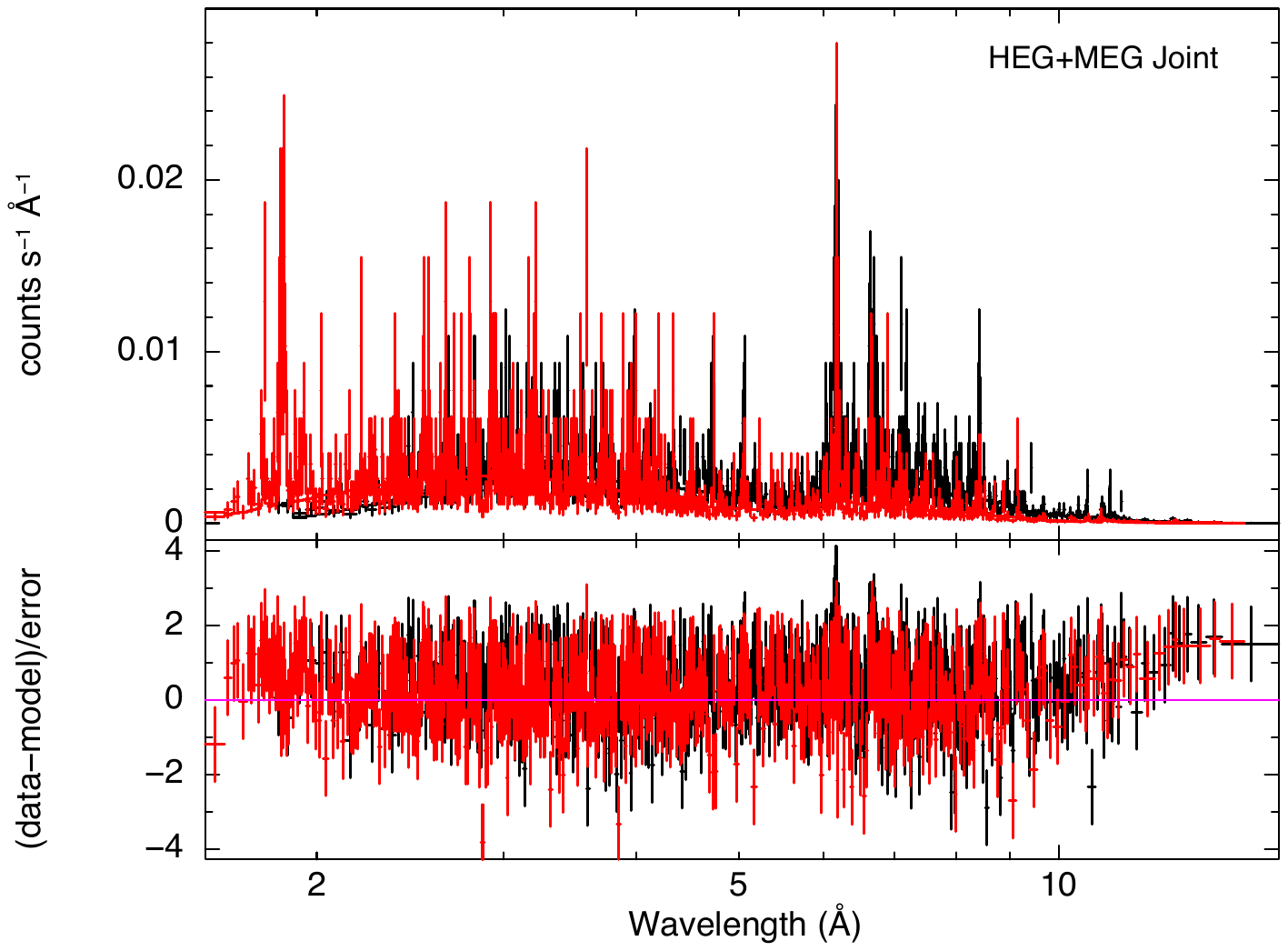}
\hspace{-0.7cm}
\includegraphics[height=7.cm,width=9.7cm,angle=0]{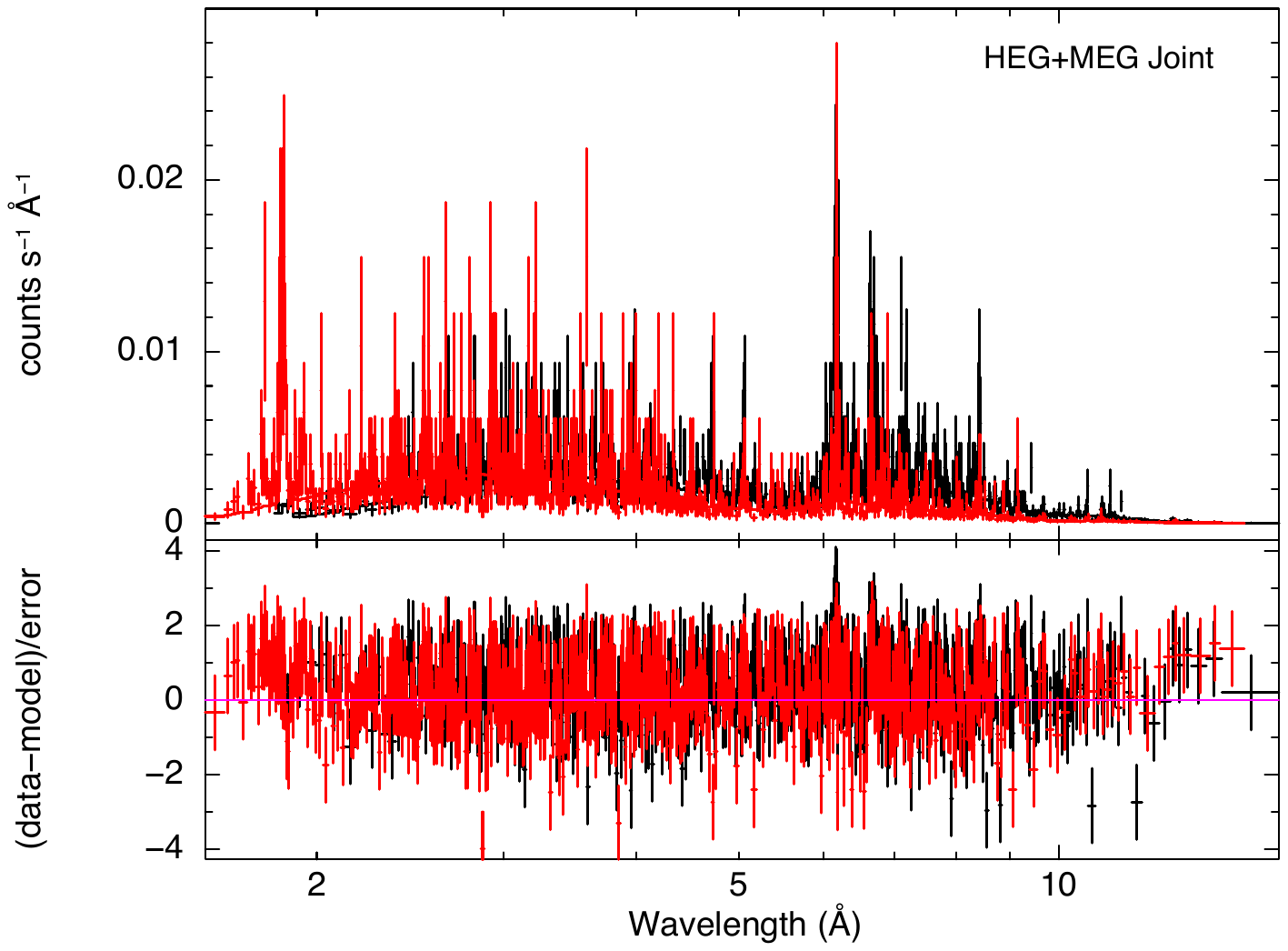}
\caption{
Fits to the joint HETG, HEG and MEG spectra of BZ Cam. HETG spectra are fit simultaneously with the  model, {\scriptsize{$tbabs$$\times zxipcf$$\times$(VNEI+power)} } on the left and {\scriptsize{$tbabs$$\times pcfabs$$\times zxipcf$(VNEI)}} model on the right. The residuals of the fits are displayed in the lower panels in standard deviations (in sigmas). \label{fig:hetgsp}} 
\end{figure*}

The basic analysis steps  of the \cha\ HETG data for BZ Cam have been detailed in Sections~\ref{sec:data} and \ref{sec:lhetg}. The CIAO task {\it tgextract2} is used to obtain the different order ($\pm$) high resolution spectra and the task {\it combine$\_$grating$\_$spectra}  is used to combine the $\pm$ orders where we chose to use only order 1 spectra since the higher order data were not useful due to low signal-to-noise. We have used both the "events and "region" options for background scaling methods in {\it tgextract2} which yielded similar results. After the line detections, identification and diagnoses of temperatures and densities, we have also performed global simultaneous fits to the joint HEG$+$MEG spectra. This spectral analysis is performed within the XSPEC software environment (for references and model descriptions, see \citealt{1996Arnaud}).  For fits we have grouped spectra using the binning-type, SNR (signal-to-noise ratio), with a binning-type value of 2 for HEG and 3 for MEG (which calculates spectral bins with a minimum of 2$\sigma$ and 3$\sigma$, respectively). We have used three different composite models (HFit1, HFit2, HFit3) to fit the joint HEG$+$MEG spectra : 1) $tbabs$$\times zxipcf$$\times$(VNEI), 2) $tbabs$$\times pcfabs$$\times zxipcf \times $(VNEI), 3) $tbabs$$\times pcfabs \times zxipcf$$\times$(power+VNEI).  Table~\ref{tab:hetgsp} displays the spectral parameters for these fits describing the high resolution spectra characteristics of BZ Cam in the 0.5-9.0 keV range (1.0-15 \AA) and Figure~\ref{fig:hetgsp} shows two of these fitted spectra.

\begin{deluxetable}{lllll}
\tablewidth{0pt}
\tablecaption{Spectral Parameters of the
Fits to the Joint HETG spectra of BZ Cam (HEG and MEG simultaneous fitting).  \label{tab:hetgsp}}
\tablehead{ 
Model  & Parameter & HFit1$^{\S{1}}$ & HFit2$^{\S{2}}$ & HFit3$^{\S{3}}$ }
\startdata
tbabs & $N_H$  & 0.1 (fixed)  &  0.1 (fixed) & 0.1 (fixed)\\
       &     (10$^{22}$atoms cm$^{-2}$) &  &  &  \\
pcfabs  & N$_H$  & N/A & $3.4^{+0.9}_{-0.8}$ & $3.4^{+0.8}_{-0.7}$ \\
   &  (10$^{22}$atoms cm$^{-2}$) &  &   &  \\ 
        & cov. frac. &  N/A & $0.61^{+0.03}_{-0.03}$ &  $0.62^{+0.03}_{-0.04}$ \\      
zxipcf  & N$_H$  &  $0.6^{+0.5}_{-0.3}$ & $1.4^{+0.8}_{-0.5}$ &  $1.4^{+0.6}_{-0.8}$ \\
   &  (10$^{22}$atoms cm$^{-2}$) &  &   &  \\ 
          & log($\xi$) & $3.6^{+0.3}_{-0.1}$ &  $3.6^{+0.2}_{-0.1}$ & $3.6^{+0.2}_{-0.2}$ \\
        & cov. frac. & $0.68 <$  & $0.77 <$ &  $0.65 <$ \\         
\hline
VNEI &  kT\ (keV)    &  $27.1^{+14.0}_{-4.5} $  & $4.9^{+0.4}_{-0.3} $ & $4.1^{+0.5}_{-1.0} $ \\
      &  $\tau\ (s cm^{-3})$  & $2.6^{+0.2}_{-0.2}\times 10^{11}$ & $4.0^{+0.5}_{-0.4}\times 10^{11}$ & $4.6^{+1.1}_{-0.8}\times 10^{11}$\\
      & K$_{VNEI}$ &  $1.7^{+0.1}_{-0.1}\times 10^{-3}$  & $2.7^{+0.1}_{-0.1}\times 10^{-3}$ & $1.9^{+0.3}_{-0.5}\times 10^{-3}$ \\
\hline
Power law & Photon Index &  N/A & N/A & $1.7^{+0.2}_{-0.3}$ \\
         & K$_{power law}$ & N/A & N/A & $2.9^{+0.5}_{-1.6}\times 10^{-4}$   \\
         \hline
 & Ne/Ne$_{\odot}$     & 1.0 (fixed) & 1.0 (fixed) & $4.4^{+4.3}_{-4.3}$ \\
 & Mg/Mg$_{\odot}$     & 1.0 & 1.0  & $2.8^{+0.9}_{-0.8} $    \\
 & Si/Si$_{\odot}$     & 1.0 & 1.0   & $1.7^{+0.4}_{-0.4} $ \\
 & S/S$_{\odot}$     & 1.0 & 1.0  & $1.1^{+0.5}_{-0.4} $   \\
  & Ca/Ca$_{\odot}$     & 1.0 & 1.0  & $1.3^{+1.0}_{-0.6} $   \\
 & Fe/Fe$_{\odot}$    & 1.0  & 1.0  & $1.2^{+0.3}_{-0.2} $  \\
\hline
   & $\chi^2_{\nu} (\nu)$  & 1.0 (1158)  & 0.94 (1156) & 0.9 (1148) \\
\hline
 Flux & (10$^{-12}$erg~cm$^{-2}$s$^{-1}$)  & 3.6$^{+0.2}_{-0.4}$ & 5.0$^{+0.2}_{-0.3}$  & 5.5$^{+0.5}_{-1.7}$ \\
 Luminosity & (10$^{31}$erg~s$^{-1}$)  & 4.7$^{+0.3}_{-0.5}$ & 6.5$^{+0.3}_{-0.4}$  & 7.2$^{+0.6}_{-2.2}$ \\
\enddata  
\tablecomments{
{\bf \S{1}}-{\scriptsize{$tbabs$$\times zxipcf$$\times$(VNEI)}};
{\bf \S{2}}-{\scriptsize{$tbabs$$\times pcfabs$$\times zxipcf \times $(VNEI)}};\\
{\bf \S{3}}-{\scriptsize{$tbabs$$\times pcfabs \times zxipcf$$\times$(power+VNEI)}};
{\it {tbabs}}--(abund=wilm)\citet{2000Wilms}. Solar abundances are assumed in the plasma models when necessary, HFit3 is performed with free abundances as relevant to the line diagnosis analysis.
$N_H$ is the absorbing column, $K_{powerlaw}$ is the normalization (i.e., photon flux in phot. cm$^{-2}$\ s$^{-1}$) for the power law model. While calculating flux and luminosity this parameter is fixed at the best fit value.
NEIvers.3.0.9 plasma code with ATOMDB database was assumed for VNEI fits.
All errors are calculated at the 90$\%$ confidence limit for a single parameter.
The unabsorbed X-ray flux and the luminosities are given in the range 0.5-9.0 keV.
The distance of 374 pc is used for luminosity calculations.}
\end{deluxetable}

In the fits we have used a $tbabs$ neutral absorber model \citep{2000Wilms},  and fixed this at the interstellar hydrogen column density value of 1.0$\times$10$^{21}$atoms cm$^{-2}$ where a conventional value is N$_{\rm H}$ in a range (0.8–1.2)$\times$10$^{21}$atoms cm$^{-2}$  derived in the broadband fits and explained in the original paper by \citet{2022Balman} (also consistent with the neutral hydrogen column density tool {\it nhtot} by \citealt{2013Willingale}) . Next, we applied one cold (neutral) partially covering absorber model, $pcfabs$, and an ionized warm absorber model that may also be partially covering, $zxipcf$. Along with these absorbers, we have used a standard nonequilibrium ionization plasma emission model, VNEI, in XSPEC. This choice is in accordance with the results of the previous spectral analysis by us in the broadband 0.1-78.0 keV range \citep{2022Balman} where find that collisional ionization equilibrium models are inconsistent with the spectra, particularly in the 6-7 keV iron line complex, together with the line diagnosis results we achieved in Section~\ref{sec:lhetg} of this paper (which are similar to our previous findings). In the last composite model HFit3 we also add to this combination a power law model of emission.
 
Our spectral results reveal that all three fits are consistent with the data so the statistics do not allow a significant choice between the composite models except that the fits HFit2 and HFit3 are inline with \citet{2022Balman}. Moreover, we recover cold and ionized absorbers using the simultaneous HETG fits. We find  a cold absorber of (2.6-4.3)$\times$10$^{22}$atoms cm$^{-2}$ (covering fraction of 0.58-0.65)  and an ionized absorber with (0.3-2.2)$\times$10$^{22}$atoms cm$^{-2}$ (covering fraction of 0.65$<$).  The plasma temperatures obtained from the nonequilibrium ionization plasma model VNEI are in a range 3.1-5.3 keV where the ionization timescale $\tau$ = (3.6-5.7)$\times$10$^{11}$\ s~cm$^{-3}$. This ionization timescale is similar to ones derived in \citet{2022Balman} and indicate that X-ray emitting plasma is in a nonequilibrium ionization state. The very high temperature of 27 keV found from HFit1 is inconsistent with previous results and the results obtained in Sections ~\ref{sec:lhetg} and ~\ref{sec:Bjoint} of this work, however can not be ruled out given statistics. We measure a photon index of 1.4-2.0 for a possible  power law model of emission in the system.  In our HFit3, we investigated non-solar abundances of  Ne, Mg, S, Si, Ca, and Fe. The statistical errors of abundances are large, but all are consistent with solar abundances and some over abundance between 2-4.

\subsection{Broadband Spectral analysis using\ \rosat, \cha, and \nustar}\label{sec:Bjoint}

A broadband spectral analysis of BZ Cam was presented in \citet{2022Balman} using the \rosat, \swi, and \nustar\ data in the 0.1-78.0 keV range, and the results were given in Table 1 of that paper. Their analyses use composite models of a cold absorber $tbabs$, $VNEI$ plasma model, BREMSS Bremssstrahlung emission model, GABS, line absorption and GAUSS, Gaussian emission line models within the XSPEC software environment. In this paper, we double checked this broadband spectrum (0.1-78 keV) with similar composite models using a spectrum derived from the zero order \cha\ data with a better spectral resolution than the \swi\ spectrum that was used to cover the medium energy ranges. 

A joint fit using \rosat, \cha\ zero order, and \nustar\ was performed in this study to check the consistency of the source spectral results, investigate absorber effects and 
possibly detect continuum components like a power law not significantly recovered in the previous paper. \rosat\ PSPC data were used to accommodate for the
softest X-ray energies in conjunction with the data from the other two
observatories. The same \rosat\ PSPC spectrum was used for this analysis, and also with a grouping of 10–20 counts in each bin to achieve adequate statistics \citep[see][for details of \rosat\ analysis]{2014Balman}. The \nustar\ observatory data were reduced as described in Section 2 and 3 of \citet{2022Balman} and source and background
spectra, response, and ancillary files were generated using the {\it nuproducts} tool for the NuSTAR data analysis. Subsequently, the FPMA/B spectra were combined and grouped with a S/N ratio of 10 in each spectral bin for good \chisq\ statistics.

The new spectrum derived from the zero-order \cha\ ACIS-S data was prepared along the standard analysis guidelines by first extracting the source$+$background, and the background spectra using CIAO task {\it specextract}  for the four individual observations where this task also produces appropriate ancillary response functions (arf files) and the energy redistribution matrix files (rmf files). A circular photon extraction radius of 6$^{\prime\prime}$ was used for the source and background events where the background spectra were extracted from a source-free zone close to the source and normalized to the same extraction area. Next, all four sets of products were combined using the script  {\it combine$\_$spectra} which produced a merged single spectrum from all four data sets where the script also
calculates a combined single response matrix and ancillary response function file. This combined spectrum was grouped with a minimum of 30 counts in a single bin to achieve good \chisq\ statistics.

\begin{deluxetable}{llll}
\tablewidth{0pt}
\tablecaption{Spectral Parameters of the
Fits to the Joint Broadband (\rosat, \cha\ zero order, and \nustar) Spectra of BZ Cam.  \label{tab:bb-sp}}
\tablehead{ 
Model  & Parameter & BFit1$^{\S{1}}$ & BFit2$^{\S{2}}$ }
\startdata
$tbabs$  & N$_H$ (atoms cm$^{-2}$) & $0.27^{+0.08}_{-0.07}\times 10^{22}$   & $0.24^{+0.06}_{-0.06}\times 10^{22}$  \\
$zxipcf$  & N$_H$ (atoms cm$^{-2}$) & $21.8^{+2.5}_{-2.2}\times 10^{22}$ & $64.4^{+13.7}_{-17.9}\times 10^{22}$ \\
          & log($\xi$) & $0.3^{+0.3}_{-0.2}$ &  $2.7^{+0.2}_{-0.3}$ \\
        & cov. frac. &   $0.67^{+0.03}_{-0.04}$ &  $0.35^{+0.06}_{-0.06}$  \\         
\hline
VNEI &  kT\ (keV)    &   N/A & $6.3^{+0.3}_{-0.3} $  \\
      &  $\tau\ (s\ cm^{-3})$  &  N/A  & $7.4^{+4.6}_{-2.1}\times 10^{11}$   \\
      & K$_{Vnei}$ & N/A &  $2.9^{+0.1}_{-0.1}\times 10^{-3}$  \\
\hline
BREMSS    & kT\ (keV) &  $ 3.4^{+0.3}_{-0.2} $ & N/A   \\
    & $K_{Bremss}$  & $2.35^{+0.07}_{-0.07}\times 10^{-3}$ & N/A   \\
\hline
{\it Power-law} & Photon Index &  $1.68^{+0.05}_{-0.05}$ & $1.0^{+0.1}_{-0.1}$ \\
         & $K_{power law}$ &  $2.3^{+0.2}_{-0.2}\times 10^{-4}$ & $1.3^{+0.4}_{-0.4}\times 10^{-5}$  \\
\hline
GABS & LineE (keV) &  1.5$^{+0.06}_{-\infty}$   & 1.4$^{+\infty}_{-0.05}$ \\
     & $\sigma$ (keV) & 0.31$^{+0.06}_{-0.05}$    & 0.34$^{+0.08}_{-0.06}$ \\
     & depth &  0.5$^{+0.1}_{-0.1}$    &  0.4$^{+0.1}_{-0.1}$  \\
GABS & LineE (keV) & 2.20$^{+0.03}_{-0.02}$    &  2.21$^{+0.02}_{-0.03}$ \\
     & $\sigma$ (keV) & 0.03$^{+0.04}_{-0.02}$   & 0.005$^{+0.002}_{-0.002}$ \\
     & depth & 0.10$^{+0.04}_{+0.03}$    &  1.8$<$ \\
GABS & LineE (keV) & 7.02$^{+0.13}_{-\infty}$    &  6.90$^{+0.18}_{-\infty}$ \\
     & $\sigma$ (keV) & 0.01 (fixed)   & 0.006$^{+0.01}_{-\infty}$ \\
     & depth & 0.09$^{+2.3}_{-0.08}$    &  0.08$^{+7.0}_{-\infty}$ \\
GAUSS & LineE (keV) & 6.65$^{+0.02}_{-0.01}$   & N/A\\
     & $\sigma$ (keV) &  0.01 (fixed) & N/A  \\
     & $K_{Gauss}$ & $2.2^{+0.3}_{-0.2}\times 10^{-5}$     &  N/A  \\
GAUSS & LineE (keV) & 6.43$^{+0.05}_{-0.05}$  & 6.45$^{+\infty}_{-0.02}$   \\
     & $\sigma$ (keV)   & 0.12$^{+0.07}_{-0.05}$   & 0.12$^{+0.05}_{-0.04}$ \\
     & $K_{Gauss}$ &  9.0$^{+0.2}_{-0.2}\times 10^{-6}$  & 1.2$^{+0.2}_{-0.2}\times 10^{-5}$ \\
GAUSS & LineE (keV) &  6.91$^{+0.05}_{-0.04}$  & 6.92$^{+0.04}_{-0.05}$     \\
     & $\sigma$ (keV)  &  0.01 (fixed)  & 0.03$^{+0.23}_{-\infty}$   \\
     & $K_{Gauss}$ &   5.7$^{+0.2}_{-0.2}\times 10^{-6}$ &   5.0$^{+2.3}_{-2.3}\times 10^{-6}$   \\
\hline
   & $\chi^2_{\nu} (\nu)$  & 1.26 (224)  & 1.27 (218)  \\
\hline
 Flux (total) & (10$^{-12}$erg~cm$^{-2}$s$^{-1}$)  & 13.6$^{+0.6}_{-0.4}$ & 8.2$^{+1.0}_{-0.5}$   \\
Luminosity (total) & (10$^{32}$erg~s$^{-1}$)  & 2.2$^{+0.2}_{-0.1}$ &  1.3$^{+0.2}_{-0.1}$  \\
 Flux (thermal)  & (10$^{-12}$erg~cm$^{-2}$s$^{-1}$)  & 9.7-11.1 &  7.2-7.4  \\
 Luminosity (thermal) &   (10$^{32}$erg~s$^{-1}$)  &  1.5-1.8 &  1.1-1.2 \\
Flux (nonthermal) & (10$^{-12}$erg~cm$^{-2}$s$^{-1}$)  &  2.8-4.0   &   0.54-1.8    \\
Luminosity (nonthermal) &  (10$^{32}$erg~s$^{-1}$)  &  0.44-0.64  &  0.1-0.3  \\
\enddata  
\tablecomments{ 
{\bf \S{1}}-{\scriptsize $tbabs$$\times$$zxipcf$$\times$GABS$\times$GABS$\times$GABS$\times$(BREMSS+power+GAUSS+GAUSS+GAUSS)};\\
{\bf \S{2}}-{\scriptsize $tbabs$$\times$$zxipcf$$\times$GABS$\times$GABS$\times$GABS$\times$(VNEI+power+GAUSS+GAUSS)}; 
{\it tbabs} model set at "abund=wilm"\ \citet{2000Wilms}. Solar abundances are assumed.
Fits are performed using \rosat, \cha\ zero order, and \nustar\ spectra in the 0.2-75.0 keV range. A multiplicative constant value model is fitted as a free parameter to account for the normalization between different missions/detectors.  $K_{powerlaw}$ is the normalization of the power law model (i.e., photon flux in phot. cm$^{-2}$\ s$^{-1}$). NEIvers.3.0.9 plasma code with ATOMDB database was assumed for VNEI fits. All errors are calculated at the 90$\%$ confidence limit for a single parameter. The unabsorbed X-ray flux and the luminosities are in the range 0.1-78.0 keV. The distance of 374 pc is used.}
\end{deluxetable}

\begin{figure*}
\includegraphics[height=5.8cm,width=6.6cm,angle=0]{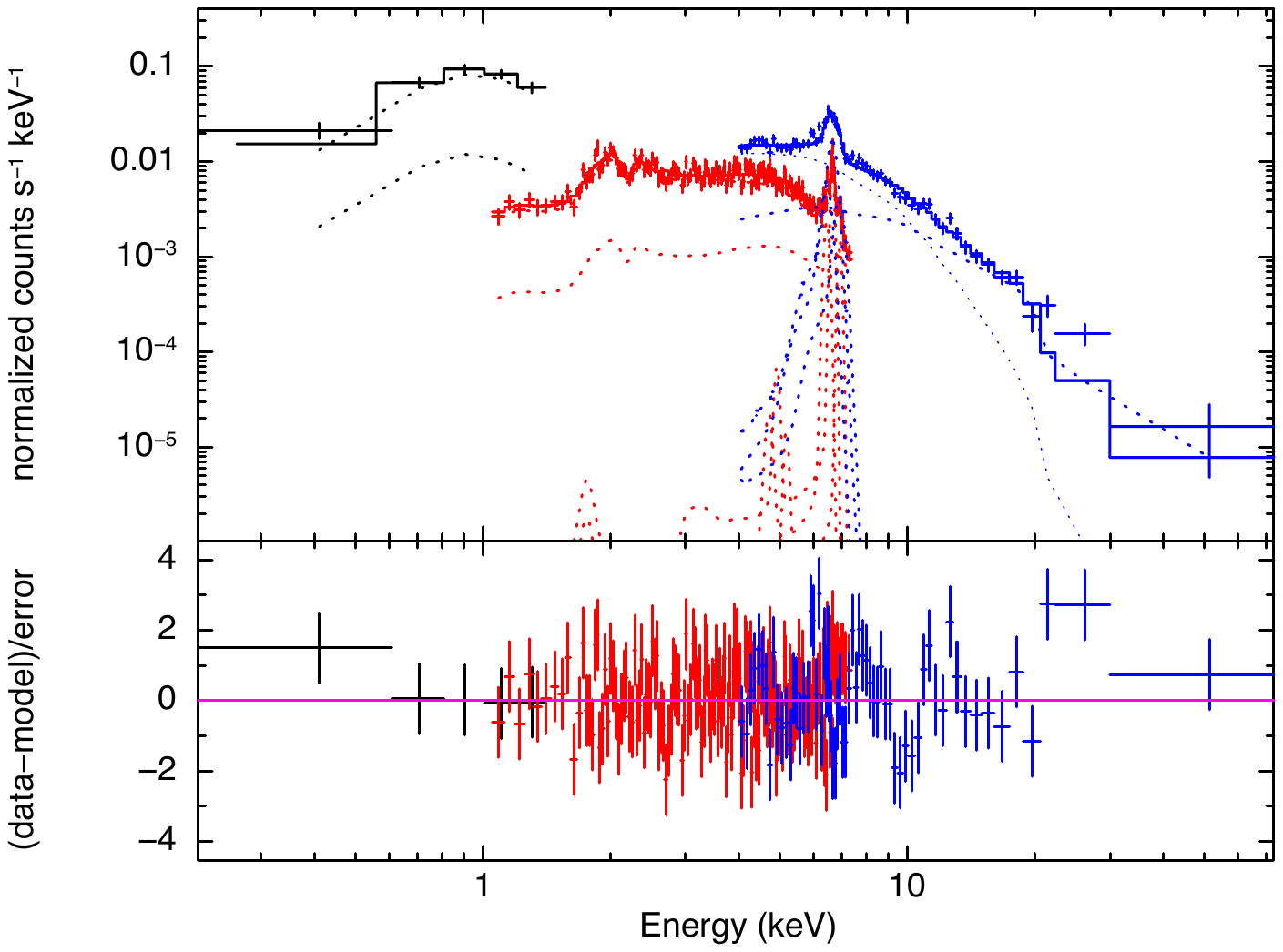}
\hspace{-0.8cm}
\includegraphics[height=5.8cm,width=6.6cm,angle=0]{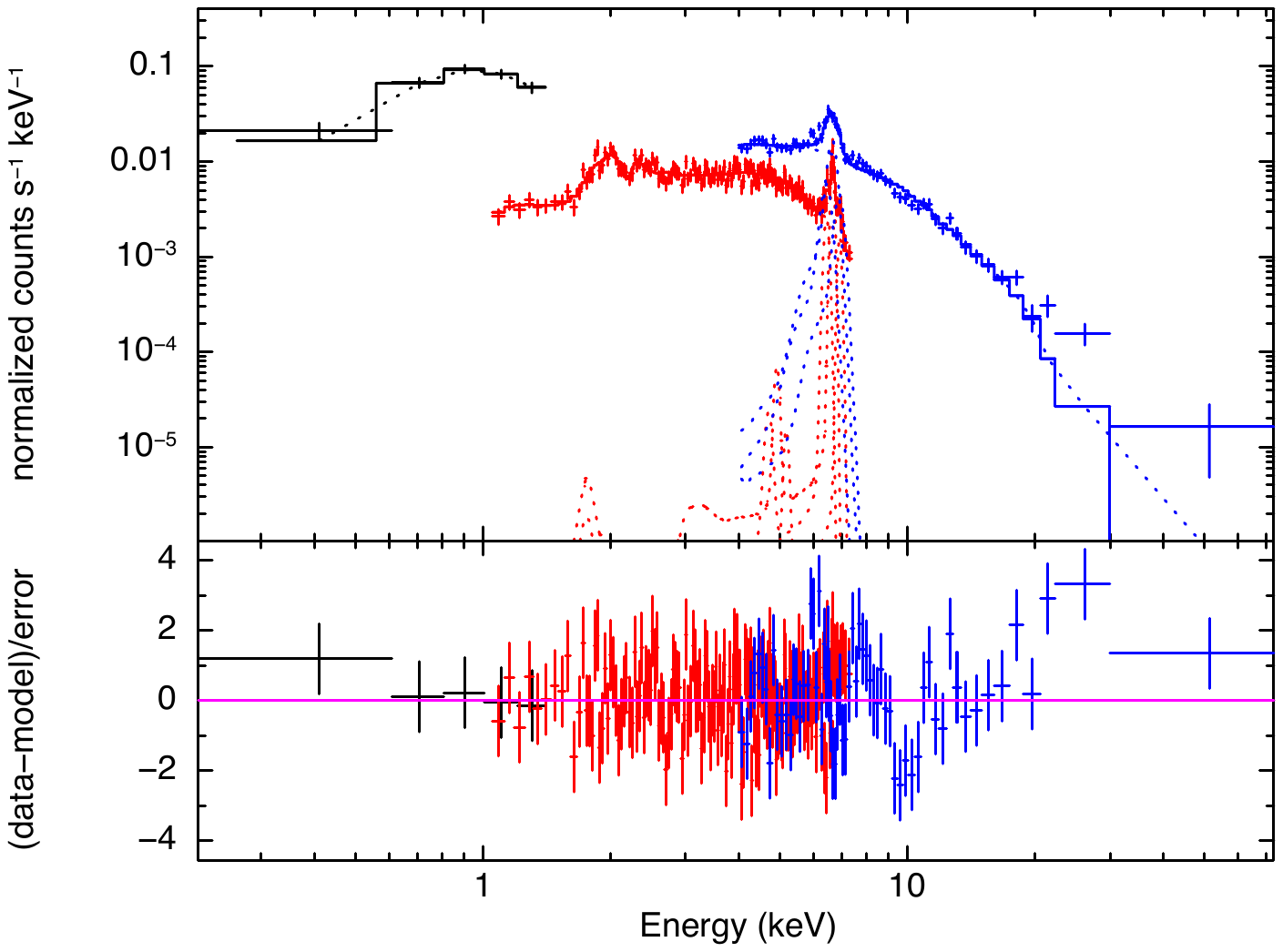}
\hspace{-0.8cm}
\includegraphics[height=5.8cm,width=6.6cm,angle=0]{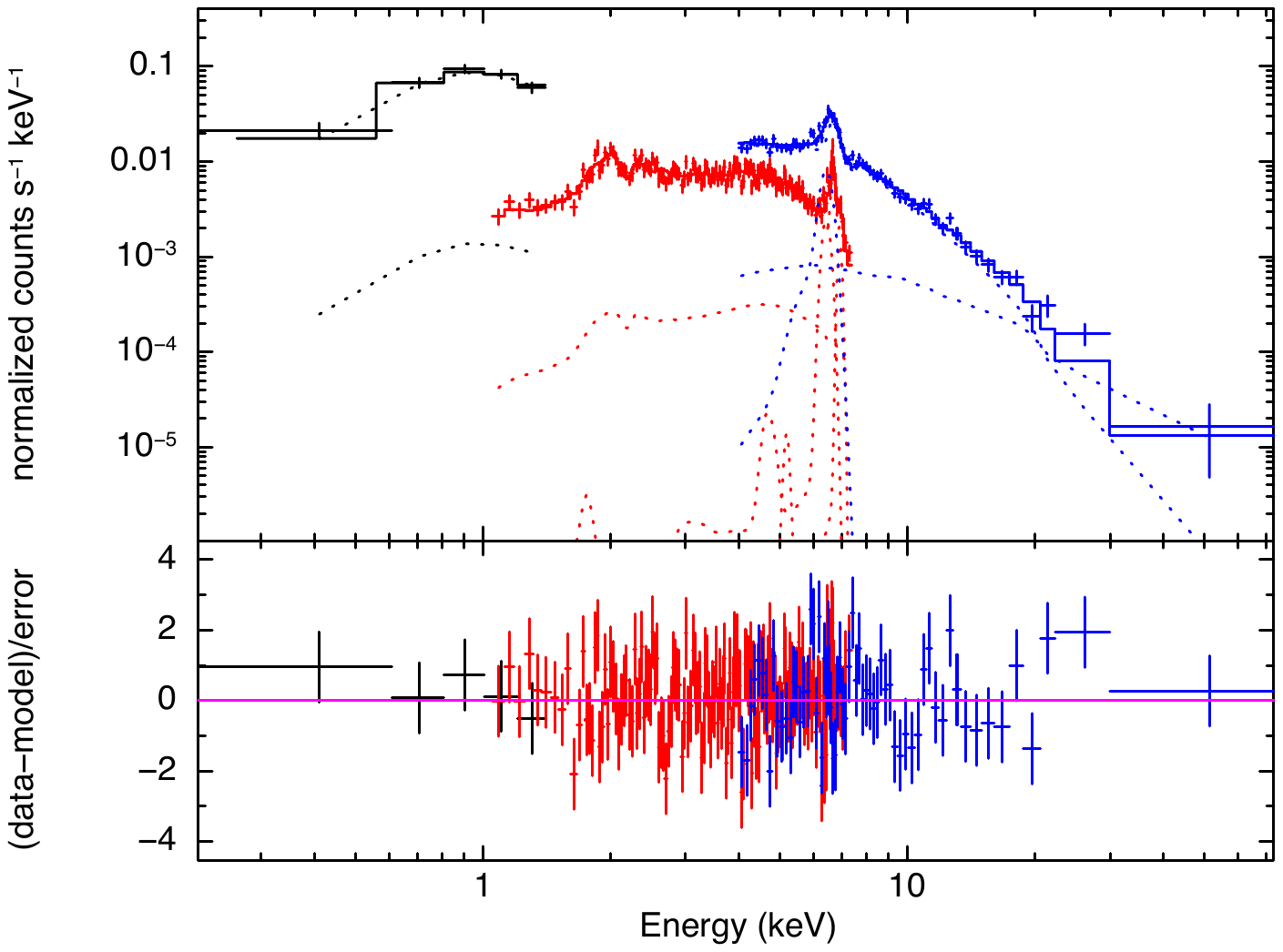}
\caption{
Fits to the joint \rosat, \cha\ zero order, and \nustar\ spectra of BZ Cam. All three spectra are fit simultaneously with the  model, {\scriptsize $tabs$$\times$$zxipcf$$\times$GABS$\times$GABS$\times$GABS(BREMSS+power+GAUSS+GAUSS+GAUSS)}
on the left. The middle panel is the same fitted joint spectra 
without the inclusion of {\it power law} model (notice the excess beyond 10 keV). The right hand panel shows the
same spectra using the composite model {\scriptsize $tbabs$$\times$$zxipcf$$\times$GABS$\times$GABS$\times$GABS(VNEI+power+GAUSS+GAUSS)}. The 
dotted lines show the contribution of different model components where the residuals  in standard deviations (in sigmas) are displayed in the lower panels. \label{fig:bbsp}} 
\end{figure*} 

The broadband spectrum of BZ Cam is fit with similar models to the previous broadband analysis using \swi\ data.  There are two composite model fits applied to the joint \rosat, \cha, and \nustar\ spectra. The first model is {\scriptsize $tbabs$$\times$$zxipcf$$\times$GABS$\times$GABS$\times$GABS$\times$(BREMSS+power+GAUSS+GAUSS+GAUSS)} (similar to the third fit in the previous paper) and the second model is {\scriptsize $tbabs$$\times$$zxipcf$$\times$GABS$\times$GABS$\times$GABS$\times$(VNEI+power+GAUSS+GAUSS)} (similar to the second fit in the previous paper).  The results of the fitting procedure are given in Table~\ref{tab:bb-sp} and example fitted models are displayed in Figure~\ref{fig:bbsp}. In this version of the modeling, we have added a partially-covering ionized absorber to the fits relying on the fact that \cha\ zero-order data has better energy resolution than \swi and better photon statistics (long exposure). In this analysis, compared with Section~\ref{sec:jhetg} we also set the $tbabs$ model free to account for the cold absorption. The first fit uses BREMSS continuum  emission model of XSPEC along with GAUSS model to account for the major emission lines and GABS model for any absorption features. The second fit utilizes the VNEI XSPEC model for a nonequilibrium ionization plasma emission with GABS model to account for the absorption and GAUSS model for the emission lines that may not be in the plasma model itself. 

The Bremsstrahlung temperature is found in a range 3.2-3.7 keV and the VNEI model plasma temperature is 6.0-6.6 keV where both ranges are in accordance with the fits using the HETG joint spectra (see Section~\ref{sec:jhetg}). The ionization timescale is also in agreement with the higher end of the range detected from HETG spectral analysis. We have used an ionized absorber (partially covering) to account for the continuum variations and if a global model requires such an absorber for a better fit. 

The two different composite models find different ionized absorbers with different  ionization parameters and covering fractions, but the same cold absorber. We tested the significance of the $zxipcf$ model by setting model parameters to zero and refitting the composite model to the spectra to see the effect on both BFit1, and BFit2. We used an F-TEST  within XSPEC that compares the \chisq\ values and degrees of freedom (dof) of these two different ﬁts. The F-TEST probability (1--conﬁdence level) was 0.0017 for BFit2 , which yields an improvement at 99.8\% Confidence Level that is  just at 3$\sigma$. On the contrary, the BFit1 with Bremsstrahlung model yields no improvement when the $zxipcf$ absorption model is removed. Moreover, we tested the significance of the additional power law model using an F-TEST which gave a probability of 0.0014 for BFit1 (i.e., BREMSS) and an improvement right around 3$\sigma$,
but the improvement for BFit2 (i.e., VNEI) was only at 98\% Confidence Level below 3$\sigma$. Note that the value for the photon index for this fit is also too flat for X-ray Binaries (XRB), in general (i.e., not a very physical additional model). The GABS models find absorption features at 1.4-1.5 keV, which may be attributed to Fe L-shell line absorption around Fe XX-XXIII, and 2.2 keV which is likely Si XII or Si XIII. We cannot measure any blueshifts since such values are within our statistical errors. The third GABS is consistent with the 6.9-7.2 keV FeXXVI H-like absorption line. 

\subsubsection{Temporal Analysis of the \cha\ data and Orbital Variations}\label{sec:lc}

In order to investigate temporal variations relevant to our analysis,  we created background-subtracted light curves with the aid of the
CIAO task {\it dmextract} to search for time variability and/or periodicity. We utilized the zero-order data with the same region files used  in the broadband analysis. A PDS calculated from the \nustar\ time series and the break frequency of 2.5$\pm$0.5 mHz can be found in \citet{2022Balman}.  We have not recovered any significant signal using  \cha\ PDS, but  the PDS  is variable at about 1\% rms level and indicates a break  f$_{br}$$<$3.5 mHz (consistent with \nustar\ analysis).  Next, we searched variability over the binary period. We used the orbital period and the ephemeris derived from the He I  $\lambda$5876 line  T$_0$ = 2453654.008(2)+0.15353(4) × E \citep{2013Honeycutt}. The error in the period of this ephemeris accumulates phase errors over one phase thus a correct phase-locking is not available. However, the phases we calculated for each of the four \cha\ light curves from each OBSID folded at this spectroscopic period are phase-locked to each other, and thus can be compared. 

\begin{figure}
\begin{center}
\vspace{-0.3cm}
\includegraphics[width=8.8cm,height=5.0cm]{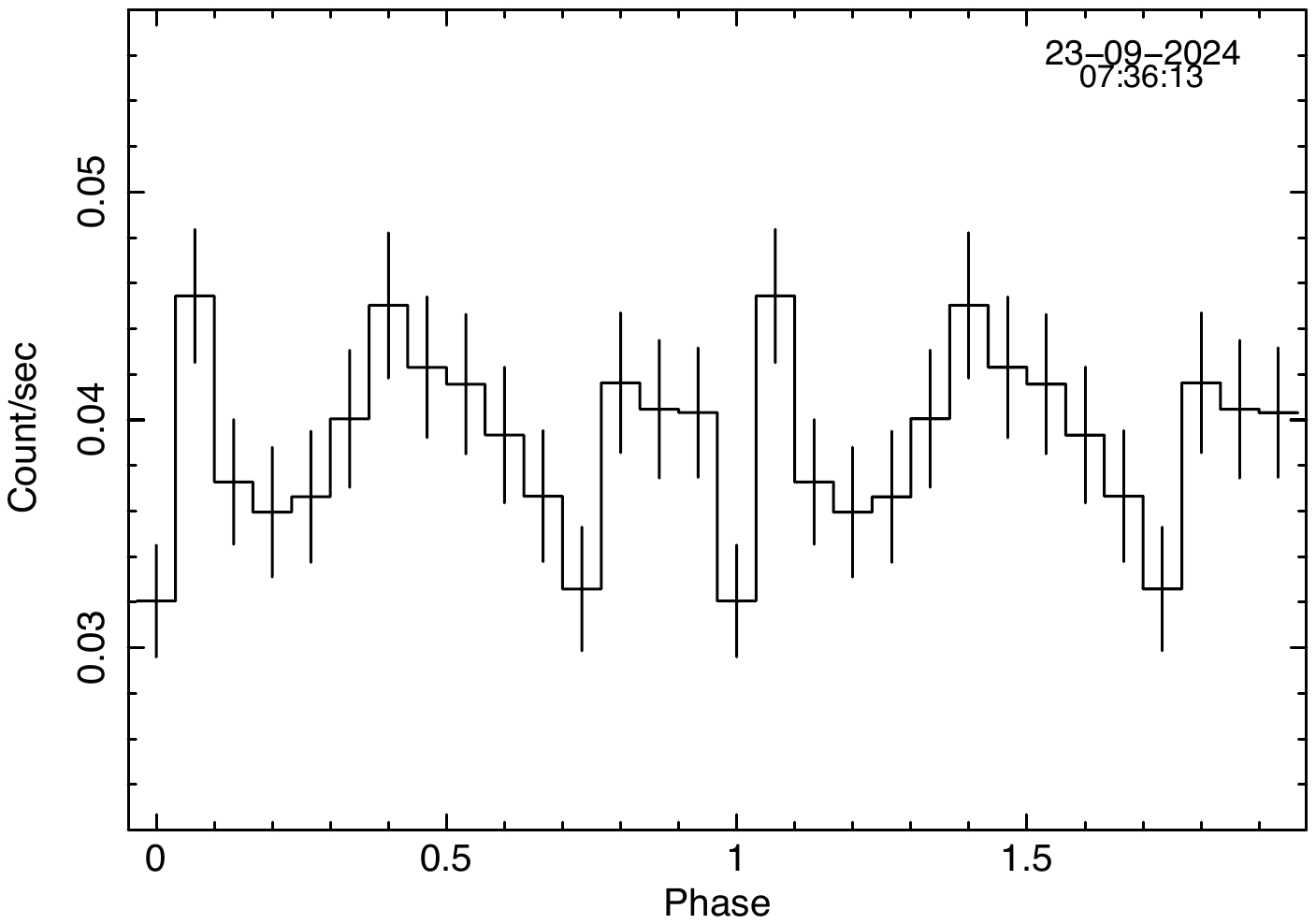} 

\vspace{-1.05cm}
\includegraphics[width=8.8cm,height=5.0cm]{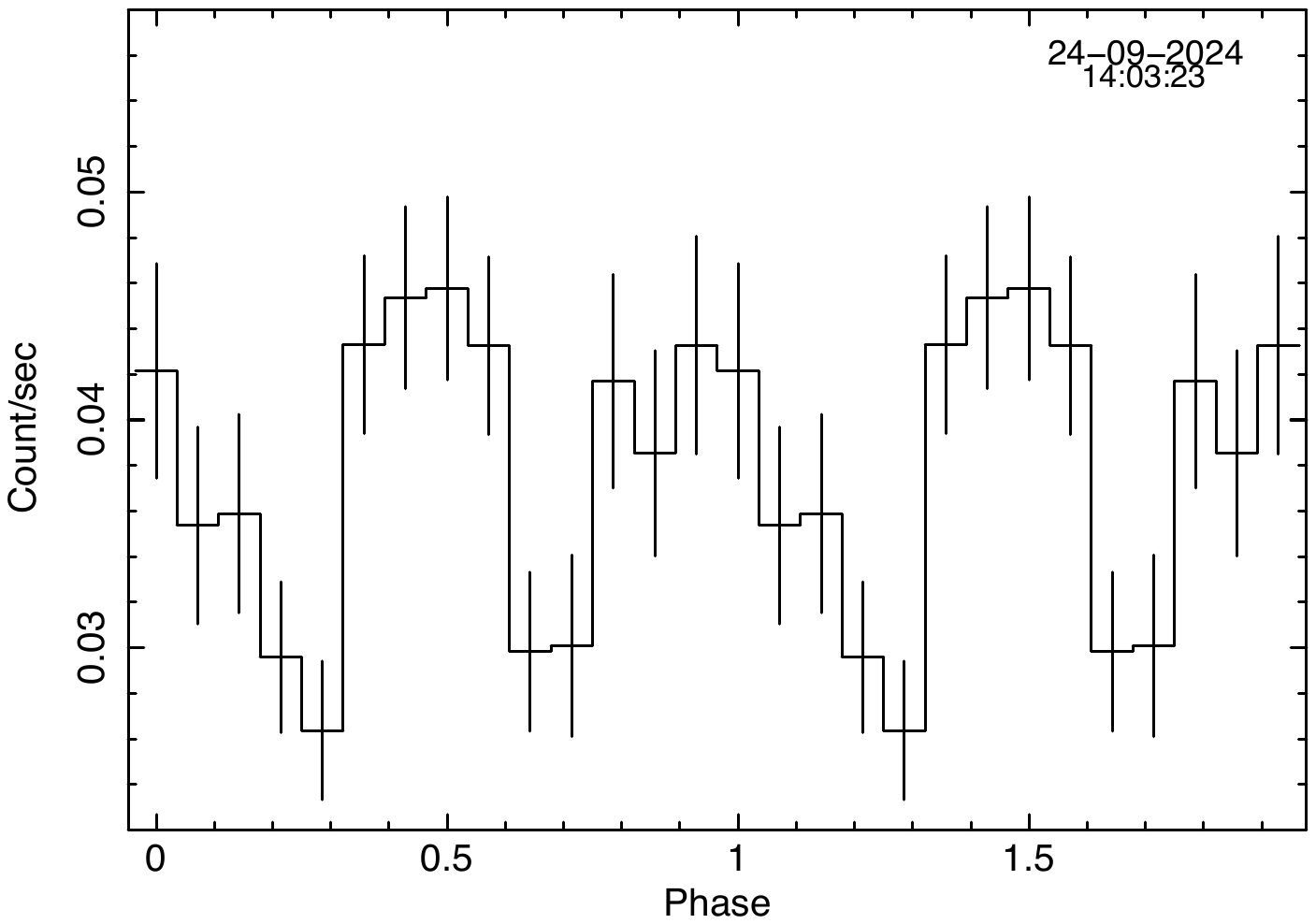} 

\vspace{-1.05cm}
 \includegraphics[width=8.8cm,height=5.0cm]{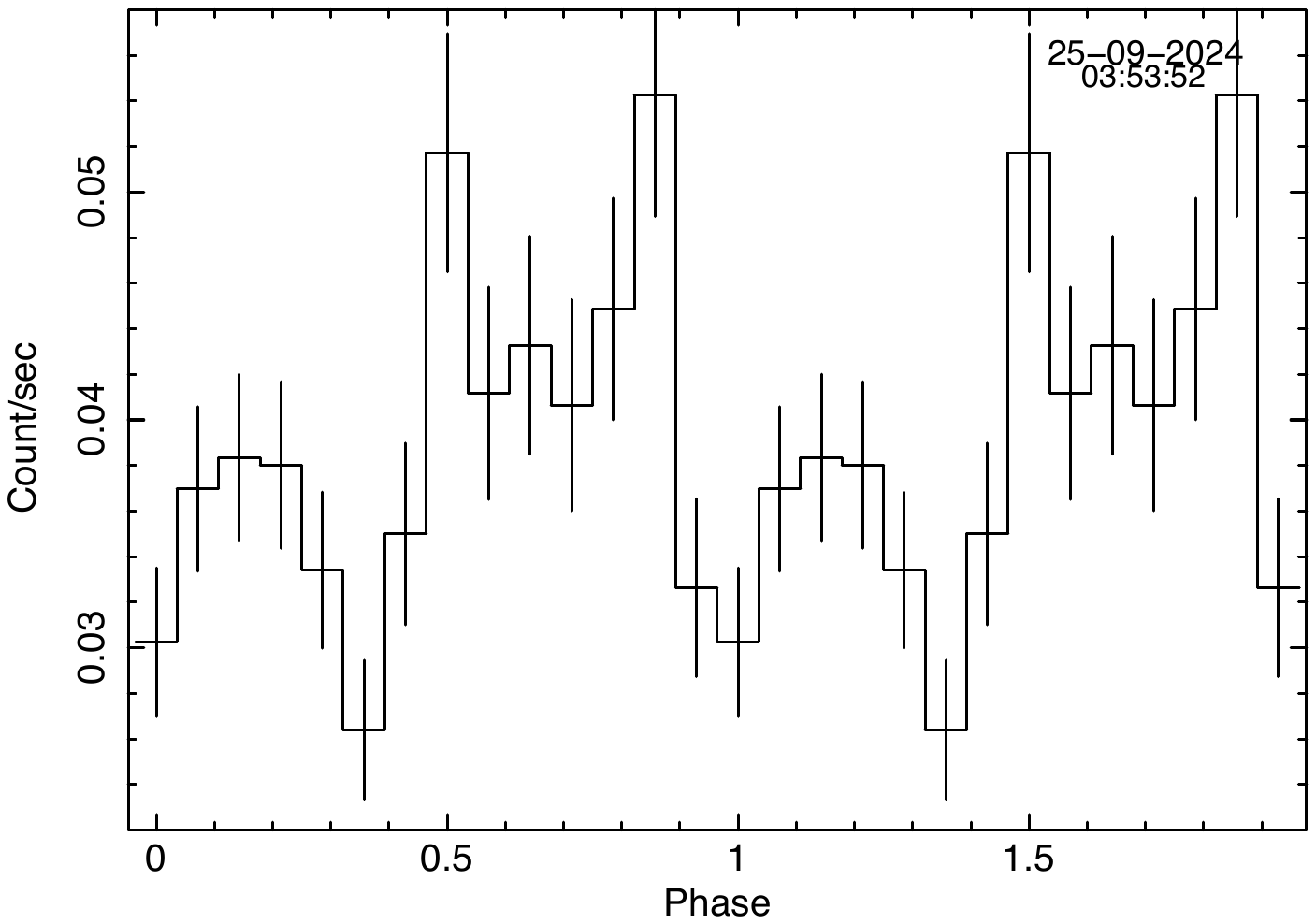} 

\vspace{-1.05cm}
 \includegraphics[width=8.8cm,height=5.0cm]{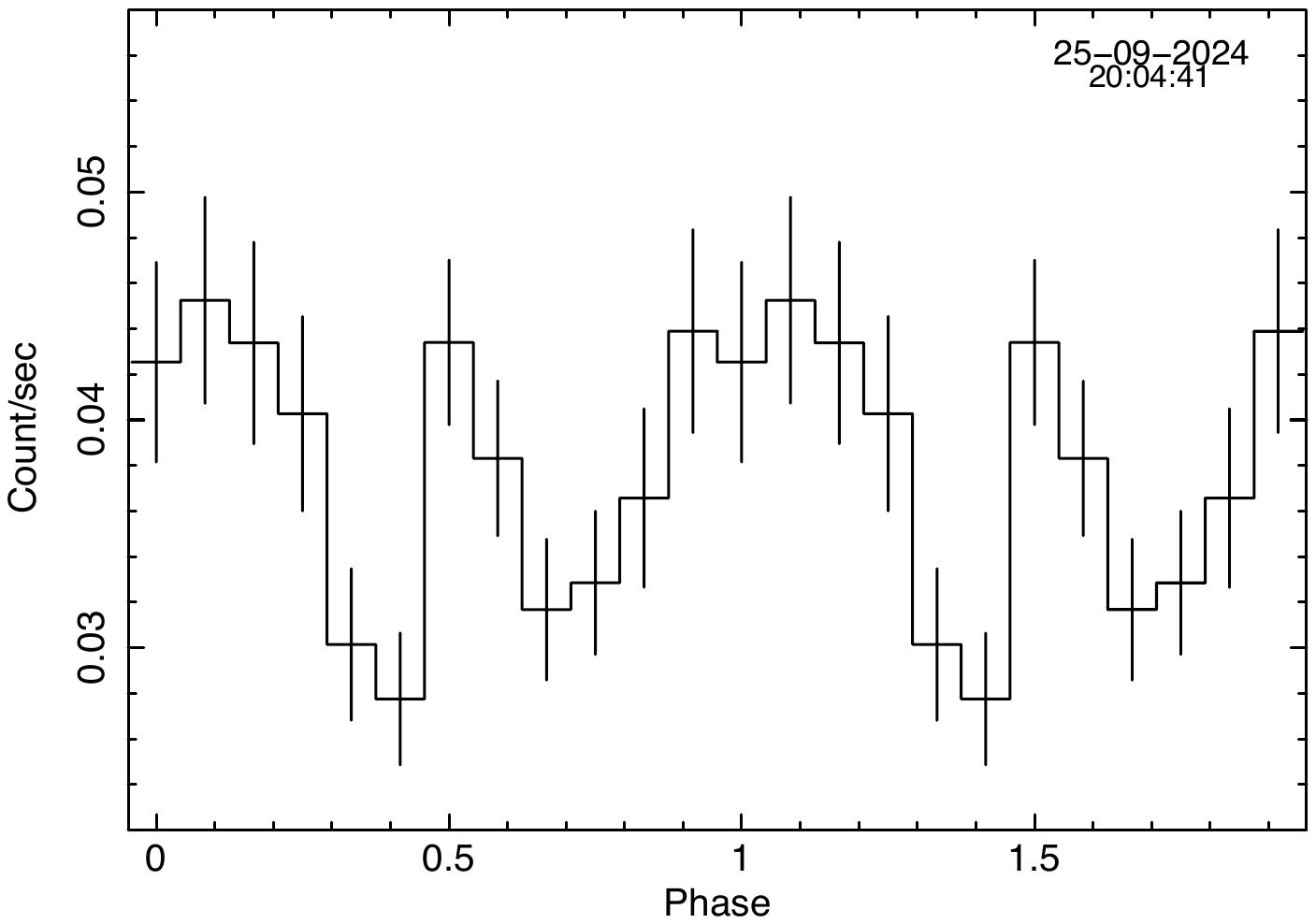} 
 
\vspace{-0.4cm}
    \caption{\cha\ zero order light curves (LCs) folded over the binary period ephemeris in Section~\ref{sec:lc} (\cha\  2024 Sep 23-25). The observation times/dates are labeled on the panels. The light curves are background subtracted and folded over the optical spectroscopic period.  \label{cha:zerolc}}
   \end{center}
\end{figure}

Figure~\ref{cha:zerolc}\ shows the four sets of light curves folded over the binary period of BZ Cam. From top to the bottom panel of the figure, the profiles are constructed over, 5.22, 2.46, 2.47, and 2.47 cycles of the period.  We calculated percent pulsed fractions over the orbital phase from an average  maximum and an average minimum of the mean light curves using $PF=[(F_{max}-F_{min})/(F_{max}+F_{min})]\times 100$.  We find for the longest (60 ks exposure) a pulsed fraction of 16\% and the other 30 ks observations show 22\%-25\% pulsed fraction where the higher variations are from the 2024 September 25 early morning observation (third panel in Figure~\ref{cha:zerolc}). These folded profiles indicate obvious dipping in the X-ray emissions/light curves. These dips change from one period to another and even if they appear in nearly similar phases, their depth and shape is variable. Thus, folding the time series over longer periods of time smears the dipping activity (top panel). As a result, we find general dipping activity over the binary period and the X-rays seem to be veiled by the material in the winds and/or from an extended disk structure which would be responsible for the ionized and cold absorbers in the system.

 In order to test this, we attempted phase-resolved spectroscopy to see if dipping is the result of absorption. We used most relevant phases of 0.25-0.45 and built spectra of the source using phased-event files. Note that we used slightly varying phases to account for the changing location of the dip structure in each light curve. A zero order spectra is extracted for each OBSID and then the resulting spectra are co-added using the task {\it combine$\_$spectra}. The phases are calculated from the "time" columns of event files using {\it dmtcalc}, then good time intervals are calculated selecting over the phases with {\it dmgti}, next the event files are filtered using the "gti" files, and finally {\it specextract} is used with relevant region files and source spectra (zero order) are calculated at the selected phases. We omitted the OBSID=28503 (third panel from the top) which yields statistically the worst spectrum. The co-added spectra yielded 430 photons, which we used to simultaneously fit with the combined average zero order spectrum of BZ Cam. During the fitting, we fixed the model to BFit2 in Table~\ref{tab:bb-sp}. We only varied the ionized absorption and cold absorption components of the model ({\it zxipcf, tbabs}). The resulting fit has a \rchisq\ of 1.3. For the cold absorption we obtained the same result within errors as in the Table for both spectra. The log($\xi$) parameter was fixed at the best fit value for the combined average spectrum and the ionized absorber parameters were varied for the phase-resolved spectrum. Finally, we found a different colder ionized absorber for the dip-spectrum as we have expected. The parameters are (at 90\% Conf. Lev.), an equivalent N${\rm_H}$ of (11.7-32.8)$\times$10$^{22}$ cm$^{-2}$, and log($\xi$)$<$1.4 (at about 60\% covering fraction). We also note that Fe abundance of the dip-spectrum is twice that of the average spectrum.

\subsection{Deriving the Mass Accretion Rate from the HST COS Data}\label{sec:uv}

\begin{figure}
\begin{center}
\includegraphics[scale=0.50,trim= 0 0 0 0.cm]{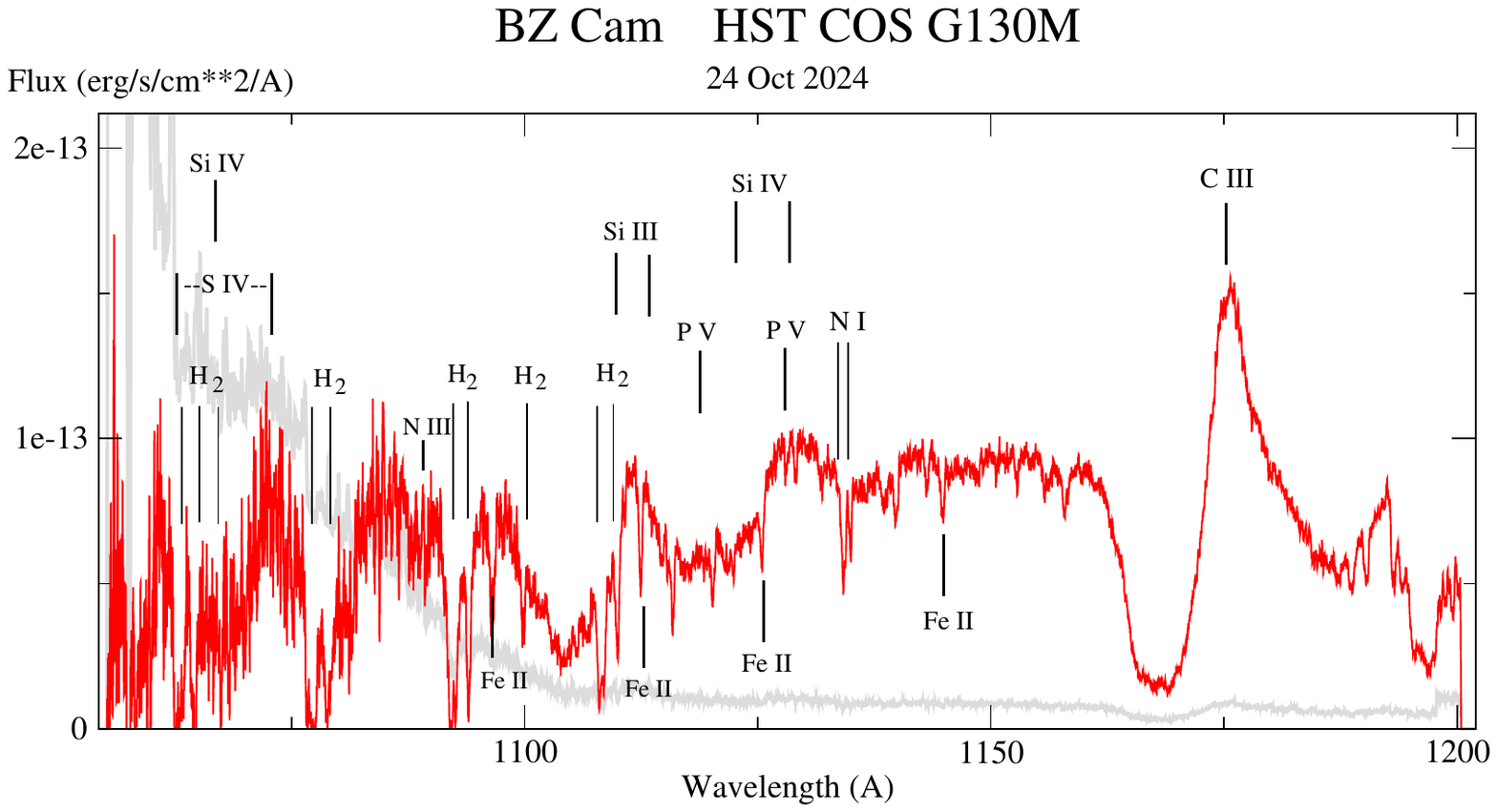}
\vspace{-2.0cm}
\caption{
The 2024 HST COS G130M (1055~\AA ) spectrum of BZ Cam with line identifications.
The spectrum is in red, for convenience the error is in grey.
The spectrum has not been dereddened.
All the narrow absorption lines are from the ISM, mainly hydrogen molecular lines
($H_2$) and iron lines (Fe\,{\sc ii}). The C\,{\sc iii} (1175~\AA ) line exhibit a prominent P-Cygni profile. We have also annotated the position of higher ionization species usually observed in disk-dominated CVs from S, Si, N, and P, but they cannot be identified in this spectrum.
\label{g130m}}
\end{center}
\end{figure}

\begin{figure}
\begin{center}
\includegraphics[scale=0.50,trim=0 0 0 +0.cm]{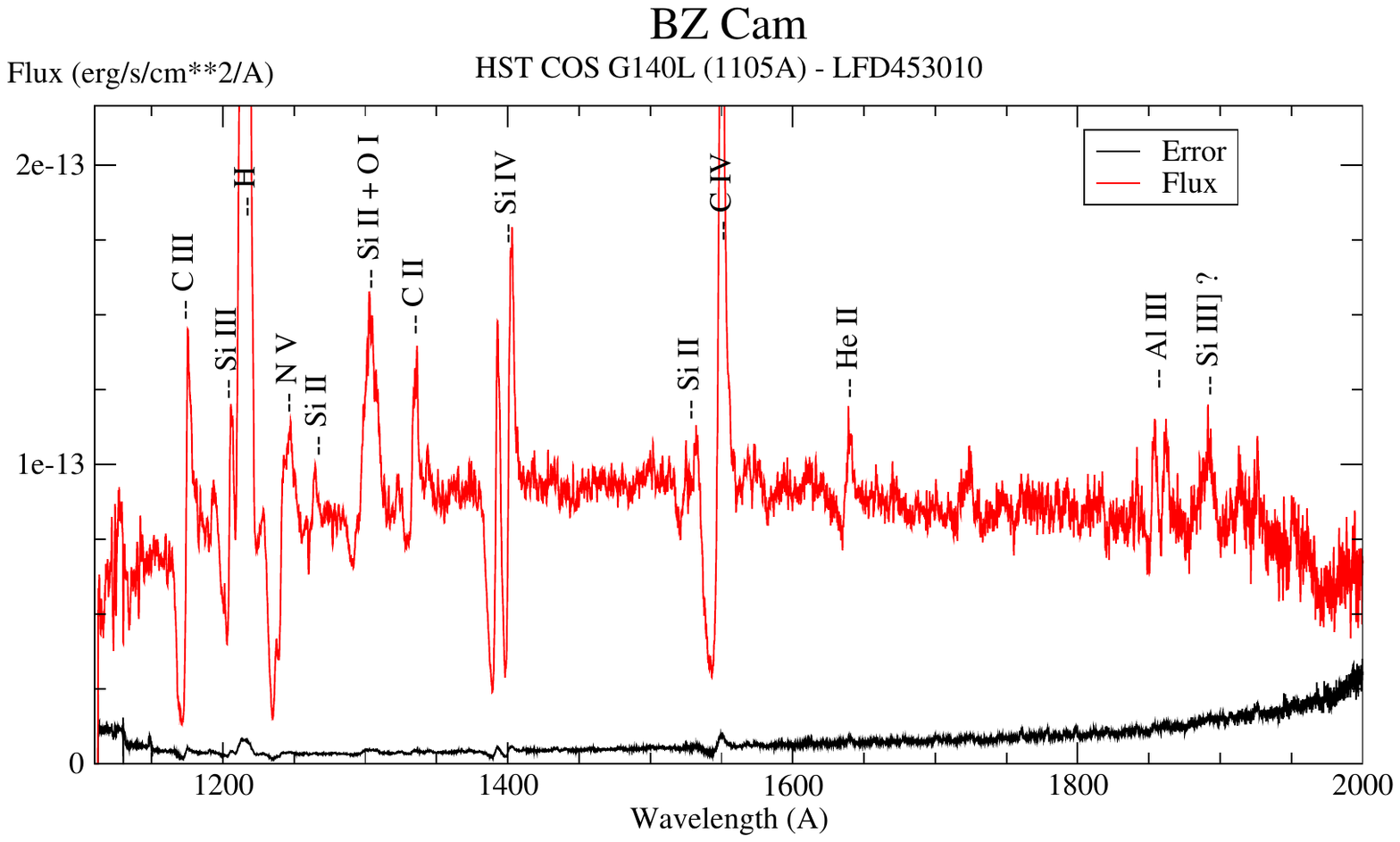}
\vspace{-1.0cm}
\caption{Line identifications in the September 2025 COS Spectrum of BZ Cam.
This spectrum (in red) has a continuum flux level about 20\% lower than the one analyzed in \citealt{2017Godon}
The error is shown in black. The spectrum has not been dereddened.
        \label{g140l}}
        \end{center}
\end{figure}

Our original goal was to carry out a multi-wavelength campaign, and for that purpose we were
able to secure HST DD (ID \# 17695) observations of BZ Cam to be carried out concurrently with the Chandra
X-ray observations. The instrumental setup was made to cover a broad wavelength region
(from $\sim 920$~\AA\ to $\sim 2150$~\AA)
while still maintaining a relatively good S/N. Approval was obtained for one HST visit using the
COS instrument set up with the G130M (1055~\AA) grating (first orbit), followed immediately by the COS
instrument switching to the G140L (1105~\AA) grating (second orbit).
The first COS (G130M) observation was carried out successfully on 2024-09-24, starting at 13:17:49 (UT)
and collecting 2308s of good exposure time (data ID LFD401020).
The second COS observation (G140L set up) failed and was re-scheduled several times.
Data (LFD453010) was eventually collected a year later on 2025-09-15 (01:06:04 UT) with 1972s of good exposure time.
Consequently, concurrent with the X-ray data,
we obtained only minimal FUV coverage (from $\approx 1000$~\AA\ to 1200~\AA , due to extremely low S/N below
1100~\AA \; see Fig.\ref{g130m}).
On the positive side, we note that on 2024 Sep. the first COS observation started about 45 min before
and ended just 20 seconds after the start of the Chandra observation.
Both COS observations lasted only about 45 min, which is a very small
fraction ($\approx 0.2$) of the binary orbital period of $\approx 3.7$~hr.
The two COS spectra are presented in Figs.\ref{g130m} \& \ref{g140l} with line identifications.

To assess the mass accretion from the UV data, we considered the first COS observations
obtained almost simultaneously with the X-ray data and scaled the second spectrum to its continuum
flux level in the region where they overlap.
We dereddened the spectrum assuming $E(B-V)=0.05$ and followed the procedure given in \citealt{2017Godon}.
The Gaia distance to BZ Cam is $374$~pc, which is a factor of 2.22 shorter
than assumed in \citealt{2017Godon}. Therefore, in \citealt{2017Godon}, when scaling the flux level
to the distance, we overestimated the mass accretion rate by a factor of 4.93.
Furthermore, the COS spectrum obtained on 24 September 2024 (at the same time as the Chandra data) has a flux slightly lower than the spectrum modeled in \citealt{2017Godon}.
Based on the 2024 Sep 24 FUV flux level and the Gaia distance we assess a (corrected)
mass accretion rate between
$3.8\times10^{-10} M_\odot$ yr$^{-1}$ and $2.1\times10^{-9} M_\odot$ yr$^{-1}$, this gives a
disk luminosity of $(2.23 \pm 0.57)\times10^{33}$ \lumcgs, with a corresponding
observable disk luminosity of $L=(1.115 \pm 0.285)\times10^{33}$ \lumcgs . This assessment is for a WD mass of $0.6 \pm 0.2M_\odot$,
with a temperature $30,000 \pm 10,000$~K, and an inclination of $i=30^\circ \pm10^\circ$.
We note that in \citealt{2017Godon} for a $0.7 M_\odot$ we had obtained that WD radius had
to be twice larger than expected, which appears to be due to the wrong distance we adopted
back then.
Therefore, the X-ray luminosity derived in the present work has to be compared to
a UV disk luminosity of $\approx (1.1 \pm 0.3)\times10^{33}$ \lumcgs.

\section{Discussion} \label{sec:disc}

\subsection{Line Diagnosis of HETG and the Nature of the Accretion Flows}
\label{sec:disc_hetg_diags}

Line ratios are utilized as diagnostics for electron density and electronic temperature.  Collisional excitation forms lines from the ground level with or without recombination.  At high temperatures (e.g., X-ray regime) the plasma can be dominated by collisional excitation or by radiative recombination, revealing photoionized plasma. Hybrid plasmas show both recombination and collisional processes. Thus, ratios of H-like and He-like ions of elements can be a probe for  collisional plasma diagnostics. The He-like triplet  is a transition between the n=2 and n=1 ground-state levels and has three components: 1) the resonance line denoted with $r$ (1s2p $^1$P$_1$ $>$ 1s$^2$ $^1$S$_0$), 2)  the intercombination line  $i$ (1s2p $^3$P$_1$ $>$ 1s$^2$ $^1$S$_0$ ) and 3) the forbidden line labeled as $f$ (1s2s $^3$S$_1$ $>$ 1s$^2$ $^1$S$_0$).  What are referred to as the ``R" and ``G" ratios between these line species 
are excellent electron density, $n_e$, and electron temperature, $T_e$, diagnostic tools, respectively \citep{1969Gabriel}.
For example, large G ratios and weak resonance lines portray photoionization and radiative recombination dominated plasmas. Instead, G $<$ 4 with dominant resonance lines would signal hybrid plasmas mediated by both photoionization and collisional processes. Collisional excitation dominates at low G ratios. 
R ratios  scale such that the ratio is high for high forbidden line intensity and low when intercombination lines are dominating the triplet emission. Low R ratio values indicate high densities and vice versa.

For BZ Cam, we have  largely detected the He-like and H-like emissions of Mg, S, Si, and Fe with some Fe-L shell emission lines. In general we find that resonance lines are not dominating and the H-like emissions are not prominent and mostly have similar or weaker intensities compared to He-like lines.  The intensity ratio of H-like emission lines to any of the three He-like lines vary between 1-2. In He-like lines we find all three components ($f, i, r$).  This is all generally consistent with collisionally-excited plasma.  The situation is different with the Fe XXV lines. These lines dominate the entire line emission and half of the intensity in line emissions come from the lines detected with HEG (see Table~\ref{tab:hlin}). The Fe XXVI emission is about 1.7 times less than the resonance line of the Fe XXV triplet, about twice the forbidden line emission and about a factor 2.5 times less than the detected intercombination line. This indicates that iron line emissions belong to a region that is more collisionally ionized, however the weaker H-like and  resonance lines signal that perhaps the culprit is photoionization and that this location is more of a hybrid plasma where also photoionization is in effect, this would boost the intercombination lines relative to the the resonance lines and we have detected that the intercombination lines are the brightest. In general, for higher densities, 1s2s $^1$S$_0$ is also depopulated to 1s2p $^1$P$_1$, suppressing the forbidden line emission in favor of intercombination line emission. However, the nonequilibrium ionization condition can reproduce such an intercombination line feature with weak resonance and forbidden line emission while the plasma is changing in ionization state/conditions from a low collisionally  ionized state  to a higher ionization level; however this need not signal a fully collisionaly ionized plasma.

We have used the He-like triplets for estimating the electron density and temperature of the X-ray emitting region/s and examine based on the R and G line ratios if the plasma is in a collisional ionization equilibrium state or not. The R and G ratios using these triplets have also been calculated in \citet{2014Schlegel} using the HETG and such diagnoses have also been performed in \citet{2024Balman} for the dwarf nova Z Cha in quiescence and outburst.  We find the\  R ratios 1) Si: 1.0$\pm$0.5, 2) S: 0.7$\pm$0.5, 3) Mg: 1.8$\pm$1.0, 4) Fe: 0.3$\pm$0.1 .  For the G ratios we get, 1) Si: 1.5$\pm$0.5, 2) S: 1.9$\pm$1.0, 3) Mg: 1.9$\pm$1.1, 4) Fe: 1.7$\pm$0.4 .  The electron densities and temperatures have been calculated following \citet{2000Porquet,2000Bautista,2014Schlegel}. The G ratios yield electron temperatures 1) Si: $\simeq$ (5-6)$\times$10$^{6}$ K, 2) S: $\simeq$ (3-4)$\times$10$^{6}$ K, 3) Mg: $\simeq$ (3-4)$\times$10$^{6}$ K, 4) Fe: $\simeq$ (1-3)$\times$10$^{7}$ K.  The electron densities follow as 1) Si: $\simeq$ 4$\times$10$^{13}$ cm$^{-3}$, 2) S: $\simeq$ (1-3)$\times$10$^{14}$ cm$^{-3}$, 3) Mg: $\simeq$ 4$\times$10$^{12}$ cm$^{-3}$, 4) Fe: $\simeq$ 10$^{16}$ cm$^{-3}$.  Note here that for the region producing iron, most likely irradiation is effective which modifies the R ratio, thus densities should be taken with care. On the other hand, one needs to consider contribution of dielectric satellite (DES) lines, particularly to the G ratio of iron K$\alpha$ complex which creates an effect of redshifted emission in the region of the complex around 6.6-6.7 keV for collisional, photoionized or hybrid plasmas of equilibrium or transient nature \citep{2001Oelgoetz,2004Oelgoetz}. Contribution from KLL (n=2) and KLn (2$<$n$\le$10) DES lines to the G ratio is considerable, thus should be called GD ratio, particularly  below the maximum abundance temperature (collisional ionization equilibrium condition). Thus, for a temperature above 4$\times$10$^{7}$ K (maximum abundance temperature), the G ratio and the GD ratio approximates. But below this, GD ratio is larger than G ratio and considerably large below  1$\times$10$^{7}$ K \citep[cf. Fig.2 \& 4, ][]{2009Oelgoetz}. Thus our intercombination Fe XXV line is at around 6.655 keV and shows this slight redshifted charactersitic and is at considerably high flux value which indicates unresolved contribution from DES. However, the G value of 1.65, we calculated, would include the effects of the DES and is actually GD. The temperature we find is below the full-ionization limit and is uneffected. Assuming the intercombination line would also be boosted by photoionization/irradiation effect, makes this Fe XXV He-like line component extremely dominant. But note here that the Fe XXV resonance line is also not dominating. We note that the resonance line has additional flux increase due to DES effects, as well. As with the G ratio, the density-sensitive R ratio is not likely to be an accurate diagnostics for Fe XXV, because the DES blends with different types of inner shell excitation and dielectric recombination satellites which may introduce temperature and ionization-state dependence in R as well. Thus, we cannot attribute density for the iron emitting region.

The G ratios are low and not over 4.0 so collisional processes occur, but resonance lines do not dominate or have relatively low flux which shows that a full collisional ionization equilibrium is not established in the plasma flow. This is aided with the relatively comparable flux in the intercombination lines and existence of appreciable  forbidden line emission in all detected triplets. The R ratios (all) indicate varying densities which signifies a possible extended patchy region with density inhomogenity. He-like iron possibly originates from a denser and compressed region (low R value), but note the DES contributions which complicates this calculation and  most likely, this Fe XXV emitting region is subjected to irradiation, as well. It seems closer to ionization equilibrium (note that this region has T$_e$ about 1-2 keV).  In general,  a plasma will reach collisional ionization equilibrium conditions when the ionization parameter  $n_e t\sim$ several$\times$10$^{13}$ s~cm$^{-3}$. For most atoms, it is possible that equilibrium can be achieved for ionization parameters a factor of 10 lower, depending on plasma temperature and electron densities and other factors that may affect the continuum characteristics \citep{2010Smith}. The ionization parameters obtained for BZ Cam using HEG and MEG spectra are in a range (2.4-5.7)$\times$10$^{11}$ s~cm$^{-3}$ (including all fits and errors).  This range indicates consistent results with the line diagnosis and yields an X-ray plasma near but not in ionization equilibrium for BZ Cam.

In general, when compared with  the HETG analysis of other CVs \citep{2014Schlegel}, the consensus of nonequilibrium ionization conditions found in that study is consistently found in this study as well, together with higher ionization in the Fe complex around 6-7 keV (Fe XXV, Fe XXVI lines).  On the other hand, we suggest that the Fe lines in the 1-2 keV regime are affected by irradiation which boosts the intercombination lines (note also additional DES contribution to {\it i, r} lines) and depopulates the forbidden line emission level. We also expect that BZ Cam, being a high state CV, should have commonalities with DN in outburst as the systems change to high state.  \citet{2024Balman} show the line diagnosis of Z Cha using \xmm\ RGS indicating that during quiescence  no resonance lines exists, with only the forbidden lines of Ne, Mg, and Si detected, together with weak H-like C, O, Ne, and Mg. The strongest line is O VIII, with $\sim$ 4$\times 10^{-14}$ erg~s$^{-1}$cm$^{-2}$. The quiescent X-ray emitting plasma is not collisional and not in ionization equilibrium, which is consistent with hot ADAF-like accretion ﬂows. The line diagnosis in the outburst shows He-like O and Ne, with intercombination lines being the strongest  along with weaker resonance lines which are particularly similar for the Fe XXV and Fe XXVI lines of BZ Cam. This indicates the plasma is more collisional and denser, but not yet in CIE, revealing ionization timescales of (0.97–1.4)$\times 10^{11}$ s~cm$^{-3}$. The ionization timescales are about 3-4 times higher in the HETG for BZ Cam (compared with Z Cha), however the total line flux excluding the Fe XXV lines is the same as Z Cha in outburst.  The X-ray luminosity though is about a factor of 20-30 times higher in BZ Cam (in a similar energy range).  Overall, both BZ Cam and Z Cha high resolution spectroscopy show  line emission consistent with narrow lines limited with the resolution of the grating instrumentation which yields sub-Keplerian rotational velocities $<$ 1000 km s$^{-1}$ for Z Cha and  for the relevant energy resolution range of  $E$/$\Delta E$=200-600 of HETG, this translates to rotational velocities $<$ 500-1500 km s$^{-1}$ which are also sub-Keplerian; a characteristics of the ADAF-like accretion flows in the X-ray emitting region.

\subsection{Comparison of HETG results and Broadband Spectral Analyses}

In general, when we compare the broadband spectral parameters of BZ Cam in Table~\ref{tab:bb-sp} with the simultaneous fits of the HEG and MEG joint spectra, the parameters are consistent.  The plasma temperatures are 3.1-5.3 keV from the HETG fits and the broad band fits yield 3.2-6.6 keV.  The total luminosities are  (5.0-7.8)$\times$10$^{31}$ erg~s$^{-1}$ using the HETG and the (1.2-2.4)$\times$10$^{32}$ erg~s$^{-1}$ using the broadband spectra. This shows only a factor of two difference in derived luminosity. Since the thermal plasma luminosities are in a range (1.1-1.8)$\times$10$^{32}$ erg~s$^{-1}$, this may be expected since HEG+MEG will miss some part of the continuum emission.  Comparing this with the earlier paper \citet{2022Balman} which includes a \swi\ data set in place of the zero order \cha\ HETG spectrum,  the plasma temperatures are 5.4-6.5 keV (similar to this paper) and the luminosity ranges are similar to the thermal luminosity (while a power law emission is included) found to be (0.74-1.02)$\times$10$^{32}$ erg~s$^{-1}$. The broadband spectrum is modeled with additional emission lines of iron at 6.65 keV and 6.91 keV, consistent with the He-like and H-like Fe. The photon-fluxes of lines in Table~\ref{tab:bb-sp}\ are in the same range with the total  photon flux of the  $f-i-r$  He-line complex, and the H-like Fe  flux in Table~\ref{tab:hlin}.

An additional power law for the broadband spectra is detected above 98\% Confidence Level and yields parameters: photon index $\Gamma$=1.63-1.73 with a luminosity of (4.4-6.4)$\times$10$^{31}$ erg~s$^{-1}$. This range  of photon index is recovered from a HEG and MEG simultaneous fit with an almost similar photon flux (i.e., normalization) which yields the same range of nonthermal luminosity. Note that the power law component in the latter fit has no additional significance when tested  (see statistics in Table~\ref{tab:hetgsp}) as opposed to the broadband fits. The existence of a power law emission model is best derived from the BFit1 (with the Bremsstrahlung emission model), yet we note that in the earlier paper \citet{2022Balman}, this was attained from the VNEI plasma model fit with a $\Gamma$=1.76-1.87 at the same photon flux level (i.e., nonthermal luminosity).  In the latter paper, a Bremsstrahlung fit did not give any significant additional power law model. Therefore, this paper yields improvement. We note that when we remove the power law model from the broadband spectral fits we get a very similar  range of plasma temperatures to those we obtained in the earlier paper, kT=8.2-9.4 keV.

The differences between the HEG$+$MEG fits and the broadband fits in this paper or the earlier paper, occur in the detected/found absorption and emission  features which is also supported with the differences in the absorption models consistent with the spectra. This may be due to the sensitivity and energy resolution of the instrumentation and may or may not be due to the state of or conditions in  the source  itself.  We note that the P Cygni profile recovered in the H-like Fe (Fe XXVI) using the \nustar\ spectrum in the earlier paper \citep{2022Balman} is not recovered from the \cha\ zero order spectrum. This is because of the sensitivity of the zero order data and more importantly the effective area that falls off sharply after the 6.65 keV (Fe XXV) line peak. Thus in the joint broadband data, the absorption line around  7.02 keV is recovered from only the \nustar\ part of the joint spectrum, but yields the same GABS (XSPEC) model parameters as in the earlier paper. We have not attempted to reveal the P Cygni profile as noted in our earlier paper. We have found that the H-like Fe line flux (using broadband spectra) is different and was found to be about 4-5 times stronger in the earlier paper than this work. In this work, it is dominated by the \cha\ zero order spectrum in the joint fitting process which signals line flux differences in differing years (2024 versus 2017).  Assuming H-like Fe is associated with the P Cygni profile and an outflow, this hints at outflow differences in flux (less in 2024).  This is consistent with the fact that the absorption line is not detectable in the \cha\ zero order spectrum (i.e., steep effective area decrease). The He-like Fe line flux (derived with the broadband analysis using \cha\ zero order) is marginally larger in this work by a factor of 1.4 times compared with the earlier paper (2017 \nustar\ data). 

Finally, we searched for  absorber characteristics given the high resolution capacity we have with HETG (HEG$+$MEG simultaneous analysis and/or zero order in the broadband joint analysis).  In the earlier paper, we were only able retrieve the interstellar absorption in the broadband spectral analysis (with \swi\ spectrum along with \rosat). In this study, during the HEG and MEG simultaneous fitting, we fixed the interstellar absorption to this value (1.0$\times$10$^{21}$ cm$^{-2}$).  We used a cold absorber ({\it pcfabs} in XSPEC) and an ionized absorber ({\it zxipcf} in XSPEC) which consistently yielded the same parameter values for all three fits (i.e., HFit1-3).  We find a cold absorber with a hydrogen column density of (4.3-2.7)$\times$10$^{22}$ cm$^{-2}$ (covering frac. 0.58-0.65) and an ionized absorber with an equivalent N$_{\rm H}$=(0.6-2.2)$\times$10$^{22}$ cm$^{-2}$  and an ionization parameter log($\xi$) of 3.4-3.8 (covering frac. $>$0.65). For the broadband spectral analysis, we used a similar choice of cold and ionized absorbers setting the {\it tbabs} model free to test a cold absorber and using an additional ionized absober {\it zxipcf} mode. The neutral hydrogen column densities are recovered to be the same for both set of fits and 2-3 times the interstellar value which is only 1.5 times less than the cold absorption N$_{\rm H}$  obtained with the {\it pcfabs} model for the HEG$+$MEG fits. On the other hand, the results show two different ionized absorbers for a Bremsstrahlung fit (BFit1) and the nonequilibrium ionization plasma fit (BFit2) with VNEI (see broadband spectral fits in Table~\ref{tab:bb-sp}). The ionization parameters (log($\xi$)) are 0.3 (BFit1) and 2.7 (BFit2) which depict colder and hotter ionized absorbers from the two different plasma fits with twice the covering fraction in the Bremsstrahlung fit (BFit1). On the other hand we have tested the significance of ionized absorbers and find that only  BFit2 (with VEI plasma model)  yields a 3$\sigma$ detection of the absorber component. The (log($\xi$))  value is different to that from the HEG$+$MEG fits (relatively colder). The equivalent N$_{\rm H}$ value is about 30 times (for BFit2)  the range derived from HEG$+$MEG fits in the broadband joint fitting. Note here that, adding a cold absorption model {\it pcfabs} to the broadband fits do not improve the fits, increasing the \rchisq\ values while fixing the {\it tbabs}  as in Table~\ref{tab:hetgsp}.

\section{Final Remarks and Conclusions} \label{sec:conc}

This study on BZ Cam comprises the first high-resolution spectroscopy and detailed  line diagnosis of an NL system comparatively with a broadband spectral analysis over 0.1-78.0 keV energy band. It improves our understanding of the physics of accretion flows in BZ Cam via the aid of \cha\ HETG observations (a total of 150 ks).  The line diagnosis using HEG and MEG show emission from Mg, Si, S, Ca, Ni and Fe lines with a total line flux of about 5.0 $\times$10$^{-13}$ \fluxcgs\ in the HEG range and 3.3$\times$10$^{-13}$ \fluxcgs\ in the MEG range. Mg, Si, S, and Fe show all He-like components ($f-i-r$) of emission and the H-like emission lines  along with some Fe L-shell line emission. The H-like to He-like resonance line ratios (or ratios of He-like components), the R ratios (0.6-1.8) and the G ratios (1.5-1.9) of the He-like lines show that  the plasma is in a nonequilibrium ionization state. This is supported with the ionization time scale of  (2.4-5.7)$\times$10$^{11}$ s~cm$^{-3}$  derived from the HETG analysis.  We also performed broadband spectral analysis in the 0.2-75.0 keV range using \cha (zero order), \rosat, and \nustar\ spectra, the latter two of which were presented in our earlier paper \citet{2022Balman}.  We find most of our results consistent across high resolution spectral modeling, line diagnosis and broadband spectral analysis. Except for He-like Fe, line emission is boosted most likely via irradiation (with additional DES  emission) thus, the intercombination line is dominating over the others. It is also possible that the Fe line is in a state closer to collisional ionization, however the low flux of the  resonance line  and the H-like line  indicate NEI conditions. 

For the first time, we find both cold absorption and ionized absorption (at 3$\sigma$) consistent with the broadband spectrum and the high resolution HETG spectra (HEG, MEG).  This is also an update on the spectral characteristics obtained in our earlier study. The log($\xi$)  is either in a range 3.4-3.8 (HEG and MEG simultaneous analysis) or 2.4-2.9 (using VNEI plasma model in the broadband analysis), but the equivalent hydrogen column density is about a factor of 30 times higher in the broadband analysis. It is possible that the broadband analysis provides a better modeling of  the continuum and the continuum absorption.  Note here that Figure~\ref{cha:zerolc},  which shows the  light curves folded at the orbital period, is indicative of this ionized absorber veiling over the orbital phase revealing dipping at various phases that changes from orbit to orbit.  We have shown in this work that this dipping activity is due to warm absorber effects and the ionized absorption increases due to lower  log($\xi$)$<$1.4 (colder ionized absorber in the dipping phases 0.25-0.45 over the orbit).

We do not find a P Cygni profile in H-like Fe as we have done in the earlier work for two reasons: 1) the H-like emission flux of Fe is about a factor of 4-5 times smaller in this work, 2) For  the HETG, both 1st order and zero order sensitivity is not enough, and the zero order effective area and thus the spectrum falls off very sharply after the He-like Fe line emission recovering only the emission component of the P Cygni briefly. We note that the He-like Fe line flux  in this work (using broadband analysis) is about 1.4 times larger than the one in the earlier paper (rather marginal difference). Therefore, we attribute the diminishing H-like Fe line flux (in this work) to change in the outflow characteristics and the possible decrease in the collimated outflows from the X-ray region.  

We do not find a significant change in total, thermal or suggested nonthermal luminosities derived in the earlier paper, although perhaps there is a factor of 1.5-2 times more total luminosity, (1.2-2.4)$\times$10$^{32}$ erg~s$^{-1}$ (0.1-78.0 keV), when including the new \cha\ data which is a marginal difference. The HST UV data indicates that the total disk luminosity is $(2.2 \pm0.6)\times 10^{33}$\fluxcgs{\AA}$^{-1}$ . Therefore, the radiative efficiencies derived here is about 0.05-0.08. This is more efficient than the previously derived factor of 0.005 which indicates relative emission differences between the X-rays and the UV (since in both this paper and the previous paper consistent distance was assumed for the two wavelength regimes).  On the other hand, a recent paper by \citet{2024Gilmozzi},  which calculated the luminosities and accretion rates for 42 NL systems using IUE  data and {\it Gaia} parallaxes, find a luminosity of  $(6.0 \pm0.8)$$\times 10^{33}$\lumcgs\ for BZ Cam. This is value brings the efficiency factor to around 0.019 (this work and Sec.~\ref{sec:uv}  both assume standard steady-state accretion disk). Therefore, this reveals  that as we progress  further out in the disk, the efficiency of emission is getting higher and towards the inner disk it is getting lower. This manifests existence of an advective (radiatively in efficient) disk in general  portraying radiatively inefficient nature over all wavelengths to a given degree. In basic terms, if a cool disc is in thermal contact with the hot flow then there is heat conduction between the two which can lead to evaporation of the disc at low mass accretion rates, where evaporation predominates in the inner disc closer to the compact object, giving rise to a radially truncated disc/hot inner flow geometry portraying radial dependence on advection and emission efficiency \citep{1999Liu,2000Rozanska,2007Mayer,2007Done,2014Yuan}.
Moreover, \citet{2024Gilmozzi} find the highest accretion rate in NLs (among the 42 they studied)  as 6.6$\times$10$^{-9}$\msun\ yr$^{-1}$ with L=$3\times 10^{34}$\lumcgs. Most luminosities are below 10$^{34}$\lumcgs. This poses great complexities as to high accretion rates of NLs  and NLs being high state CVs when the highest accretion state achieved can be as high as 10$^{38}$\lumcgs. This puzzle can be alleviated by advective nature of accretion (i.e., radiative inefficiency) throughout the disk.

Using the HETG analysis, we find that all detected lines are narrow and rotational velocities are sub-Keplerian with $<$ 500-1500 km s$^{-1}$ which is a characteristic of ADAF-type accretion flows. We detect a power law component at 98\% Confidence Level with  $\Gamma$=1.63-1.73 and a luminosity of (4.4-6.4)$\times$10$^{31}$ erg~s$^{-1}$ which is about 0.8\% of the disk luminosity (taken after \citealt{2024Gilmozzi}).  Finally, in this work, we find significant effects (3$\sigma$) of ionized absorbers (warm absorbers) supported with the dipping and veiling detected as superposed over the light curves folded at the orbital period. 

In this study, we support our previous conclusions on BZ Cam, where we ﬁnd that the X-ray energy spectral and power spectral characteristics reveal radiatively inefﬁcient advective hot accretion ﬂows (RIAF ADAF-like) in this  system. In this regime, below 10$^{-2}$$\dot{M}_{Edd}$ (this also depends on the square of the $\alpha$ parameter), advection is effective in the accretion ﬂow, and as the accretion rate increases, the fraction of advection in the ﬂow drops \citep[][and references therein]{2008Narayan,2014Yuan} and collisional dynamics increases (collisional equilibrium in plasma flow and/or Compton effect and power law emission). Note that the fraction of advection is a function of $r$ in the disk and the accretion rate, where the optically thick part of the disk ﬂow can be sustained. We have derived this transition radius where the flow characteristics change, using power spectral analysis and calculating a break frequency in the broadband noise ($<$ 3.5 mHz and also see Section 5.1 in \citealt{2022Balman}). The discussion of what ADAF flows are and detailed characteristics in relation to accreting WDs, have been done in the earlier paper  \citep[see Section 5.2 in][]{2022Balman} and discussions in relation to state transitions and other X-ray binaries have been presented in \citet{2024Balman} and will not be repeated in this paper. We underline that the type of ADAF flows in BZ Cam and other similar NLs or DNe are "{\it thermal ADAFs}" where contribution from nonthermal components are not existent or not as effective (in relation to other X-ray binaries), mostly being NEI-type sub-Keplerian extended plasma flows (accretion flows) in the inner disk where shocks may or may not occur.

\begin{acknowledgments}

This research  and paper is based on observations obtained with \cha\ Observatory; OBSID=28050, 28502, 28503, 28504. Support for this work was provided by the National Aeronautics and Space Administration through Chandra Award Numbers  GO4-25016A and GO4-25016B issued by the Chandra X-ray Center (via Guest Investigator Program Cycle 25), which is operated by the Smithsonian Astrophysical Observatory for and on behalf of the National Aeronautics Space Administration under contract NAS8-03060. This research has made use of software provided by the Chandra X-ray Center (CXC) in the application packages CIAO 4.15-4.17. This research has made use of data and/or software provided by the High Energy Astrophysics Science Archive Research Center (HEASARC), which is a service of the Astrophysics Science Division at NASA/GSFC. The UV part of this work was based on observations (GO DD 17695 Cycle 31) made with the NASA/ESA Hubble Space Telescope obtained from the Mikulski Archive for Space Telescopes at the Space Telescope Science Institute, which is operated by the Association of University for Research in Astronomy, Inc. under NASA contract NAS 5-26555. 

This research employs a list of \cha\ datasets obtained by the \cha\ Observatory, contained in~\dataset[DOI:10.25574/cdc.557]{https://doi.org/10.25574/cdc.557}.
Some of the data presented in this article, obtained from the Mikulski Archive for Space Telescopes (MAST) at the Space Telescope Science Institute, can be accessed via~\dataset[DOI:10.17909/7apa-tn73]{https://doi.org/10.17909/7apa-tn73}.

\end{acknowledgments}

\bibliography{cam-cas}
\bibliographystyle{aasjournalv7}

\end{document}